\documentclass[floatfix,aps,prd,showpacs,amsmath,nofootinbib,preprintnumbers]
{revtex4}

\usepackage{graphicx}
\usepackage{amssymb}

\begin{document}

\preprint{FERMILAB-PUB-04-328-A} \preprint{MIT-CTP 3565}

\title{Isocurvature constraints on gravitationally produced superheavy
dark matter}

\author{Daniel J. H. Chung}\email{danielchung@wisc.edu}
\affiliation{Department of Physics, University of Wisconsin, Madison, WI 53706}

\author{Edward W. Kolb}\email{rocky@fnal.gov}
\affiliation{Particle Astrophysics Center, Fermi
        National Accelerator Laboratory, Batavia, Illinois \ 60510-0500, USA \\
        and Department of Astronomy and Astrophysics, Enrico Fermi Institute,
        University of Chicago, Chicago, Illinois \ 60637-1433 USA}

\author{Antonio Riotto}\email{antonio.riotto@pd.infn.it}
\affiliation{INFN, Sezione di Padova, via Marzolo 8, I-35131, Italy}

\author{Leonardo Senatore}\email{senatore@mit.edu}
\affiliation{Center for Theoretical Physics, MIT, Cambridge, MA 02139}

\date{\today}

\begin{abstract}
We show that the isocurvature perturbations imply that the
gravitationally produced superheavy dark matter must have masses
larger than few times the Hubble expansion rate at the end of
inflation. This together with the bound on tensor to scalar
contribution to the CMB induces a lower bound on the reheating
temperature for superheavy dark matter to be about $10^{7}$
GeV. Hence, if the superheavy dark matter scenario is embedded in
supergravity models with gravity mediated SUSY breaking, the gravitino
bound will squeeze this scenario. Furthermore, the CMB constraint
strengthens the statement that gravitationally produced superheavy
dark matter scenario prefers a relatively large tensor mode amplitude
if the reheating temperature must be less than $10^{9}$GeV.
\end{abstract}


\maketitle

\section{Introduction}

The primordial density perturbations which provide initial conditions
for structure formation \cite{lr} can be classified as two types:
curvature and
isocurvature. Isocurvature perturbations are defined to be
fluctuations in the composition of the energy density. Hence, the
curvature remains fixed while the relative contributions of various
fluid elements composing the total energy density changes.  More
explicitly, the isocurvature perturbation due to a cold dark matter
particle species $X$ can be written as
\begin{equation}
\delta_S \equiv \frac{\delta n_X}{n_X} -\frac{\delta n_\gamma}{n_\gamma} ,
\end{equation}
where $n_\gamma$ is the density of photons and $n_X$ is the
density of the dark matter particles.  As can be seen in the
appendix, generically, isocurvature perturbations are almost
always synonymous with entropy perturbations.  Although the
isocurvature perturbations do not give rise to potential energy
during radiation domination, after matter domination occurs, the
isocurvature perturbations also become source for gravitational
potential energy leaving an imprint on the CMB.

In the late 80's and early 90's, isocurvature and curvature
perturbation models were seen to be competing models for the
theory of initial conditions for structure formation. Today, from
the precise measurements of the CMB acoustic peaks and initial
measurements of the TE correlations
\cite{Peiris:2003ff,Page:2003fa}, most people agree that the
dominant contribution to the primordial density fluctuations arise
from curvature perturbations.  Nonetheless, as demonstrated by
many recent works
\cite{Peiris:2003ff,Moodley:2004nz,Bucher:2004an,Bucher:2000hy,Enqvist:2000hp,
Beltran:2004uv,Crotty:2003rz,Amendola:2001ni}, an order 10\%
contribution to the CMB power spectrum from isocurvature is
possible. Perhaps more importantly, we generically expect from
theoretical considerations a nonvanishing contribution to the CMB
from isocurvature perturbations. (In other words, whenever there
is more than one energy component in the universe, there is always
some amount of isocurvature perturbations.)

In the most popular WIMP CDM (cold dark matter) picture in
inflationary cosmology, the isocurvature perturbations are
expected to be small because CDM abundance today is determined
from a freeze out process starting from chemical equilibrium
initial conditions. Since the chemical equilibrium is with
radiation, which, by definition, has overdensities defined by
curvature fluctuations during radiation domination, CDM in this
case will naturally have suppressed isocurvature component since
\begin{equation}
\frac{\delta n_{X}}{n_{X}}=\frac{\delta n_{\gamma}}{n_{\gamma}}
=\frac{3\delta T}{T} ,
\end{equation}
where $X$ denotes the dark matter density, $\gamma$ denotes
radiation, and $T$ denotes the equilibrium temperature.

However, for superheavy dark matter which never equilibrates, the
isocurvature perturbations are generically not negligible.
Furthermore, because, for gravitational particle production
scenario, the superheavy dark matter \emph{homogeneous} density is
determined by the same inflationary dynamics as the isocurvature
perturbations themselves, one obtains a correlated constraint for
the superheavy dark matter abundance and the isocurvature
perturbation amplitude. In this paper, we compute the generic
feature of this correlated constraint.

The gravitational production of superheavy dark matter has been well
studied in the past
\cite{Chung:2001cb,Chung:1998zb,Kuzmin:1998uv,Kuzmin:1998kk,Chung:1998bt}. In
this scenario, the nonadiabatic change in the particle dispersion
relationship due to a relatively sudden transition from a quasi-de
Sitter phase of inflation to a small power law expansion phase of FRW
cosmology causes particle production. The only non-generic feature of
this dark matter scenario is the existence of stable particles with
heavy mass. For $10^{11}-10^{13}$GeV particles, up to dimension 8
decay operators needs to be absent to have the particle be long lived
enough to be CDM. Nonetheless, there are many particle physics and
string theory motivated candidates for superheavy dark matter (see for
example
\cite{Benakli:1998ut,Han:1998pa,Hamaguchi:1998wm,Benakli:1998sy,
Leontaris:1999ce,Hamaguchi:1999cv,Dvali:1999tq,Coriano:2001mg,
Uehara:2001wd,Shiu:2003ta,Ellis:2004cj}).  Hence, what makes this
scenario exciting is that if we can measure the signatures of
superheavy dark matter, we would have a probe of the very early
universe where the physics far beyond the standard model of particle
physics was important.  (Indeed, it may even probe string theory.)
This paper serves as one step towards identifying the signature of
superheavy dark matter on the CMB.

In particular, we investigate how much superheavy dark matter
isocurvature perturbations are produced at around 50 efolds before the
end of inflation and how those isocurvature perturbations will show up
on the CMB. We find that the mass of the superheavy dark matter cannot
be too small if one is to avoid overproducing isocurvature
perturbations inconsistent with the CMB data. More specifically, the
constraint is
\begin{equation}
\frac{m_{X}}{H_{e}}\gtrsim\mathcal{O}(5)
\label{eq:intromassbound} ,
\end{equation}
where $m_{X}$ is the mass of the superheavy dark matter and
$H_{e}$ is the Hubble expansion rate at the end of inflation. One
implication of this is to strengthen the preference of large
tensor perturbations for gravitationally produced superheavy dark
matter scenario.  More specifically, if the reheating temperature
is less than $10^{9}$ GeV, then the tensor to scalar power ratio
is bounded from below by
\begin{equation}
\frac{P_T}{P_R} \gtrsim 10^{-3} .
\label{eq:lowerboundongwave}
\end{equation}

Another implication of Eq.\ (\ref{eq:intromassbound}) is that
there is a lower bound on the reheating temperature for superheavy
dark matter scenario to be viable since the number density of
particles produced falls off as $m_{X}/H_{e}$ and increasing the
reheating temperature increases the final number density. Putting
this together with the bound on the tensor perturbations from
inflation, we find the approximate requirement
\begin{equation}
T_{RH} \gtrsim10^{7} \left(\frac{0.2}{r_{S}}\right)\textrm{GeV}
\label{eq:introreheatingbound}
\end{equation}
in order to have the superheavy dark matter be the CDM while not
contributing too much to isocurvature perturbations where $r_{S}$
represents the bound on the ratio of tensor to scalar perturbation
power. Hence, we have a severe restriction on this scenario coming
from the gravitino bound (see for example
\cite{Weinberg:1982zq,Ellis:1984er,Lindley:1986wt,Kawasaki:1986my,
Moroi:1993mb,Kohri:2001jx,Kawasaki:2004qu,Kawasaki:2004fw}) if
this scenario is embedded in gravity mediate SUSY breaking
scenario (for an introductory review of status of low energy
supersymmetry, see for example, \cite{Chung:2003fi}). Note that
given that both the superheavy dark matter and the gravity waves
are produced by inflationary energy density, it is natural that
there is a bound of the form Eq.\ (\ref{eq:introreheatingbound}).
What is not obvious about Eq.\ (\ref{eq:introreheatingbound}) is
how the bound on isocurvature perturbations make Eq.\
(\ref{eq:introreheatingbound}) independent of the mass of the
superheavy dark matter. The explanation of this is one of the main
results of this paper.

In the present work, we will consider superheavy dark matter model of
massive scalar field $X$ minimally coupled to gravity without any
other interaction for $X$.  The results are not likely to change
drastically with fermionic superheavy dark matter since none of the
physics relies upon the existence of a vacuum expectation value of the
dark matter field nor spin statistics.  Furthermore, we restrict
ourselves to minimal coupling in this paper.  For any fixed dark
matter mass, nonminimal coupling will change the total amount of dark
matter produced, the spectral distribution of the dark matter, and the
dark matter's stress tensor correlator.  The most important effect is
likely to come from the dark matter's stress tensor correlator.
Because a coupling of the form $\xi \phi^2 R$ (where $R$ is the Ricci
scalar) effectively shifts the mass of the superheavy dark matter as
$m_X^2 \rightarrow m_X^2 + 12 \xi H^2$ (where $H$ is the expansion
rate during inflation), a positive $\xi$ results will relax the mass
bound shown in Eq. (\ref{eq:intromassbound}), possibly to an extent
where there will be no constraint from isocurvature perturbations.  On
the other hand, a negative value for the nonminimal coupling will make
the bound more stringent.

The order of presentation is as follows. In Section 2, we make general
estimates to establish the results associated with Eqs.\
(\ref{eq:intromassbound}), (\ref{eq:lowerboundongwave}), and
(\ref{eq:introreheatingbound}). Following that, we review the basic
physics of CMB and its relationship to isocurvature perturbations.  We
then compute the isocurvature power spectrum (essentially stress
tensor correlation function) in Section 4. In Section 5, we present
numerical results with a special emphasis on a particular inflationary
model as a check of our general argument. Finally, we conclude in
Section 6. In the appendix A, we give an elementary thermodynamic
discussion of isocurvature perturbations to remind the reader of its
physics. The technical details of the particle production computation
for the toy model used in Section 5 is presented in Appendix B.

\section{General Arguments}

It is possible to semi-quantitatively describe the mass bound coming
from isocurvature perturbations as follows. Field fluctuations are
typically characterized by its $N$-point moments, and the most elementary
nontrivial moment for perturbative theories is typically the two point
function. The power spectrum of any two point function in a translationally
invariant space is defined as
\begin{equation}
P_{X}(k)=\frac{k^{3}}{2\pi^{2}}\int d^{3}re^{-i\vec{k}\cdot\vec{r}}
\langle X(\vec{x})X(\vec{y})\rangle ,
\end{equation}
where $\vec{r}$ is defined to be $\vec{r}\equiv\vec{x}-\vec{y}$.
As we will discuss below, the isocurvature perturbations that we
are concerned with are characterized by the power spectrum
\begin{equation}
P_{\delta_{X}}(k)\equiv\frac{k^{3}}{2\pi^{2}}
\int d^{3}re^{-i\vec{k}\cdot\vec{r}}
\langle\delta_{X}(\vec{x})\delta_{X}(\vec{y})\rangle ,
\end{equation}
where $\delta_{X}\equiv\delta\rho_{X}/\rho_{X}$
is the CDM (superheavy dark matter in our case) energy overdensity.

As an order of magnitude estimate, we can represent the energy overdensity
field as $\delta\rho_{X}\sim m_{X}^{2}X^{2}$, which gives
\begin{equation}
P_{\delta_{X}}(k) \sim \frac{k^{3}m_{X}^{4}}{2\pi^{2}\rho_{X}^{2}}
\int d^{3}r\ e^{-i\vec{k}\cdot\vec{r}}
\langle X^{2}(\vec{x})X^{2}(\vec{y})\rangle
 \sim \frac{k^{3}m_{X}^{4}}{2\pi^{2}\rho_{X}^{2}}
 \int d^{3}r\ e^{-i\vec{k}\cdot\vec{r}}
 \langle X(\vec{x})X(\vec{y})\rangle^{2}.
\end{equation}
Hence, in the long wavelength limit, we can approximate this as
\begin{equation}
P_{\delta_{X}}(k)\sim\frac{m_{X}^{4}P_{X}^{2}(k)}{\rho_{X}^{2}}.
\end{equation}

For scalar fields $X$ in FRW spacetimes in general, if $V''(X)\gg H^{2}$
(where $V$ is the scalar field potential and $H$ is the expansion
rate), we know that the correlation function will behave as in flat
Minkowski space. In that case, we know from Minkowski space field
theory that the two point correlation functions receive most of its
power from short distances. We also know that for long wavelengths
($k\rightarrow0$) in de Sitter (dS) space, when $V''\ll H^{2}$, that
\begin{equation}
P_{X}\sim \left(\frac{H}{2\pi}\right)^{2}\left(\frac{k}{aH}\right)^{j} ,
\label{eq:powerspecform}
\end{equation}
where $(k/aH)^{j}$ represents the factor that is breaking scale
invariance due to the fact that $V''(X)\neq0$.\footnote{It is
important to emphasize that the form of Eq.~(\ref{eq:powerspecform})
is only valid when $V''/H^2 \ll 9/4$ in the minimal coupling scenario.
On the other hand, for nonminimal coupling, this form is valid only if
$V''/H^2 \ll 9/4-12\xi$.} Hence, using the Minkowski intuition and the fact
that there is generically large power contribution in the infrared for
$V''<0$ (due to instability), we can estimate the power index of Eq.\
(\ref{eq:powerspecform}) as
\begin{equation}
j=Q\frac{V''}{H^{2}} ,
\end{equation}
where $Q>0$ is a positive order unity dimensionless number. We thus find
\begin{equation}
P_{\delta_{X}}(k)\sim\frac{m_{X}^{4}}{\rho_{X}^{2}}
\left(\frac{H}{2\pi}\right)^{4} \left(\frac{k}{aH}\right)^{2QV''/H^{2}}.
\end{equation}

Since inflation eventually ends and since the gravitational particle
production does not really {}``occur'' until the end of inflation,
we have to assign the time at which we evaluate the time dependent
quantities carefully. Since the isocurvature fluctuations are generated
as the scales leave the horizon early in the inflationary epoch, we
assign $H$ to the expansion rate at about 50 efoldings before the
end of inflation. On the other hand, the scaling of the wave vector
continues until the end of inflation at which time particles are
nonadiabatically
produced. After the nonadiabatic particle production occurs, the power
spectrum should remain approximately time independent in the long
wavelength limit owing to the arguments of causality and adiabatic
evolution. Hence, we write
\begin{equation}
P_{\delta_{X}}(k)\sim
\frac{m_{X}^{4}}{\rho_{X}^{2}(t_{e})}\left(\frac{H_{I}}{2\pi}\right)^{4}
\left(\frac{k}{a_{e}H_{I}}\right)^{2QV''/H_{I}^{2}},
\label{eq:orderofmag}
\end{equation}
where $H_{I}$ represents the expansion rate at about 50 efolds before
the end of inflation and $\rho_{e}$ represents the CDM energy density
at the end of inflation. In Section 4 we will compute this more carefully
and find that although this gives the right order of magnitude, there
is comparable contribution coming from the kinetic terms neglected
in this estimate.

The dark matter energy that appears in the denominator of Eq.\
(\ref{eq:orderofmag})
is not easy to estimate for minimal coupling because significant particle
production contribution comes from the infrared which have ambiguous
vacuum boundary conditions. As shown in the appendix, neglecting this
infrared contribution, one finds
\begin{equation}
0.7\times10^{-3}m_{X}(m_{\phi}m_{X})^{3/2}
\exp\left(\frac{-5m_{X}}{m_{\phi}}\right)
\leq \rho_{X}(t_{e}) \leq 1.6\times10^{-3}m_{X}(m_{\phi}m_{X})^{3/2}
\exp\left(\frac{-4.2m_{X}}{m_{\phi}}\right)
\label{eq:quadraticinflationbound}
\end{equation}
for $V=\frac{1}{2}m_{\phi}^{2}\phi^{2}$ potential. More generally,
we expect that
\begin{equation}
\rho_{e} \sim 10^{-3}m_{X}(H_{e}m_{X})^{3/2}
\exp\left(\frac{-2m_{X}}{H_{e}}\right) .
\label{eq:generalized}
\end{equation}
Hence, we generically find
\begin{equation}
P_{\delta_{X}}(k)\sim600\frac{H_{I}}{m_{X}}\left(\frac{H_{I}}{H_{e}}\right)^{3}
\left(\frac{k}{a_{e}H_{I}}\right)^{2Qm_{X}^{2}/H_{I}^{2}}
\exp\left(\frac{4m_{X}}{H_{e}}\right) ,
\label{eq:firstestimate}
\end{equation}
where we have assumed $V''\approx m_{X}^{2}$ consistently with the
particle production. Although naively, one would guess that the exponential
would dominated on the right hand side of Eq.\ (\ref{eq:firstestimate}),
in reality the $k$-dependent factor dominates since for large scales
of interest for CMB, we have
\begin{equation}
\ln\left(\frac{k}{a_{e}H_{I}}\right) \sim
\ln\left(\frac{a_{0}H_{0}}{a_{e}H_{I}}\right)
\sim -55+\ln\left[\left(\frac{T_{RH}}{10^{9}\textrm{GeV}}\right)^{-1/3}
 \left(\frac{V_{e}}{10^{15}\textrm{GeV}}\right)^{-1/6}\right] ,
\end{equation}
which implies that the $k$-dependent term dominates:
\begin{equation}
P_{\delta_{X}}(a_{0}H_{0})\sim 600 \frac{H_{I}}{m_{X}}
\left(\frac{H_{I}}{H_{e}}\right)^{3}
\exp\left(\frac{4m_{X}}{H_{e}}-110Q\frac{m_{X}^{2}}{H_{I}^{2}}\right) .
\end{equation}
Hence, we see the inflationary model dependence of the power sepctrum
clearly in the ratio of $H_{I}/H_{e}$ (ratio of the expansion rates
at the beginning and end of inflation) and $m_{X}/H_{I}$. Despite
this model dependence, we would generically conclude $m_{X}/H_{I}>O(1)$.
More precisely, since the isocurvature contribution must not exceed
the adiabatic contribution to $C_{l}$, we have for small $l \sim 2$
\begin{equation}
C_{l=2}^{(X)}\sim0.1 P_{\delta_{X}}(a_{0}H_{0}) \lesssim 10^{-10} ,
\end{equation}
which gives a bound of
\begin{equation}
\frac{m_{X}}{H_{e}}\gtrsim\frac{4}{\sqrt{Q}x}\sqrt{1+\frac{.05}{Qx^{2}}}
+\frac{1}{Qx^{2}} ,
\label{eq:genmassbound}
\end{equation}
where $x\equiv7H_{e}/H_{I}$ is close to unity. If we for example
take $Q=2/3$ and $x=1$, we obtain $m_{X}/H_{e}\gtrsim6$.

Since this is similar to the result of our more careful computation,
we will use this bound as our example for the rest of this section.
However, more generally, the bound is $m_{X}/H_{e}\gtrsim\mathcal{O}(5)$.

Now, according to our estimate of particle production shown in the
appendix, the dark matter density today is given by
\begin{equation}
\Omega_{X}h^{2} \approx 4.31\times10^{-5} \frac{T_{RH}}{T_{0}}
\frac{\rho_{X}(t_{e})}{\rho_{I}(t_{e})}
\sim  10^{3}\left(\frac{m_{X}}{10^{13}\textrm{GeV}}\right)^{2}
\frac{T_{RH}}{10^{9}\textrm{GeV}}
\left(\frac{m_{X}}{H_{e}}\right)^{1/2}e^{-2m_{X}/H_{e}}.
\label{eq:relicdensity}
\end{equation}
This implies that if $m_{X}/H_{e}\gtrsim6$, we must increase the
relic abundance by increasing $m_{X}$ from $3\times10^{13}$GeV
while keeping the exponential fixed or increase $T_{RH}$ or increase
both $m_{X}$ and $T_{RH}$. Note that $H_{e}$ dependence only comes
through $m_{X}/H_{e}$.

Keeping $T_{RH}$ fixed, increasing $m_{X}$ from $10^{13}$ GeV for a
fixed ratio $m_{X}/H_{e}\gtrsim6$ is restricted by a bound on the
amplitude of the gravitational wave (limiting $H_{e}$). The CMB
temperature fluctuation normalization \cite{Peiris:2003ff} fixes
$P_{\mathcal{R}}\approx2\times10^{-9}$ and since the tensor
perturbation power spectrum is
\begin{equation}
P_{T}=\left(\frac{H_{I}}{\bar{M}_{P}(2\pi)}\right)^{2},
\end{equation}
where
$\bar{M}_{P}=2\times10^{18}$GeV is the reduced Planck mass, we have
the tensor to scalar ratio
\begin{equation}
\frac{P_{T}}{P_{\mathcal{R}}} = 5\times10^{8}
\frac{H_{I}^{2}}{\bar{M}_{P}^{2}(2\pi)^{2}} =
0.02\left(\frac{H_{I}}{10^{14}\textrm{GeV}}\right)^{2},
\label{eq:tensortoscalar}
\end{equation}
which may be a detectable amplitude for $H_{I}=10^{14}$GeV
\cite{Knox:2002pe,Song:2003ca}.  If the experimental bound on the
tensor to scalar ratio is $r_{S}$
(\textit{i.e.,} $P_{T}/P_{\mathcal{R}}<r_{S}$), we have
\begin{equation}
H_{I}<7\times10^{14}\sqrt{r_{S}}\textrm{GeV} ,
\label{eq:grwavebound}
\end{equation}
and consequently, we cannot raise $H_{e}$ much above $10^{14}$
GeV.\footnote{Eq.~(\ref{eq:grwavebound}) with $r_S=1$ is about the
same bound one obtains in single field slow roll inflationary
scenarios.} Therefore, by Eq.\ (\ref{eq:tensortoscalar}) and recalling
that we are for the moment focusing on fixing $m_{X}/H_{e}\gtrsim6$
such that $\Omega_{X}h^{2}\sim0.1$, we find the bound on the
superheavy dark matter mass of
\begin{equation}
6 \lesssim \frac{m_{X}}{H_{e}} \lesssim 12 \left[1+\frac{1}{24}\ln\left(
\frac{H_{e}^{2}}{H_{I}^{2}}r_{S}\right)+\frac{1}{24}\ln
\frac{T_{RH}}{10^{9}\textrm{GeV}}\right] ,
\end{equation}
where we have approximately solved a transcendental equation by
iteration.  This corresponds to an upper bound on the dark matter mass
of
\begin{equation}
m_{X} \lesssim 10^{15} \left(\frac{7H_{e}}{H_{I}}\right)
\sqrt{r_{S}}\left[1+\frac{1}{24}\ln\left(\frac{H_{e}^{2}}
{H_{I}^{2}}r_{S}\right)
+\frac{1}{24}\ln\frac{T_{RH}}{10^{9}\textrm{GeV}}\right].
\label{eq:massupperbound}
\end{equation}

Now, suppose we want to find the minimum allowed $T_{RH}$ while maintaining
$m_{X}/H_{e}\gtrsim6$. We can then solve for $T_{RH}$ in Eq.\
(\ref{eq:relicdensity})
with $\Omega_{X}h^{2}=0.1$ and use Eq.\ (\ref{eq:grwavebound}) to
obtain
\begin{equation}
T_{RH} \gtrsim 10^{7}\frac{0.2}{r_{S}}\left(\frac{H_{I}}{7H_{e}}\right)^{2}
\textrm{GeV},
\label{eq:reheatingbound}
\end{equation}
where the $0.2$ factor would be much smaller if we were not restricted
to $m_{X}/H_{e}\gtrsim6$. This may be dangerous for the gravitino
bound if the scenario is embedded in gravity mediated SUSY breaking
scenario. In our explicit computations for $V(\phi)=\frac{1}{2}
m_{\phi}^{2}\phi^{2}$
model of slow roll inflation, we obtain a bound of $T_{RH}\gtrsim10^{8}$
GeV.

From Eqs.\ (\ref{eq:massupperbound}) and (\ref{eq:reheatingbound}),
we see that if we tighten the bound on the tensor amplitudes (by making
$r_{S}$ smaller), the superheavy dark matter scenario can be severely
restricted. In particular, the linear dependence of Eq.\
(\ref{eq:reheatingbound}) on $1/r_{S}$ can change the reheating lower bound
significantly.

Note if we require that $T_{RH} \lesssim 10^{9}\textrm{GeV}$
(for example, because we want to evade the gravitino bound), we have
from Eq.\ (\ref{eq:relicdensity}) and Eq.\ (\ref{eq:genmassbound}),
that
\begin{equation}
\Omega_{X}h^{2} \lesssim 10^{-2}
\left(\frac{m_{X}}{10^{13}\textrm{GeV}}\right)^{2}.
\end{equation}
Since $\Omega_{X}h^{2}\gtrsim0.1$, we conclude $m_{X}\gtrsim3\times10^{13}$
GeV, which implies
\begin{equation}
H_{e}\gtrsim3\times10^{12}\textrm{GeV} ,
\label{eq:helowerbound}
\end{equation}
or according to Eq.\ (\ref{eq:tensortoscalar}) that
\begin{equation}
\frac{P_{T}}{P_{\mathcal{R}}} \gtrsim
10^{-3}\left(\frac{H_{I}}{7H_{e}}\right)^{2}.
\label{eq:grwavebound2}
\end{equation}
Contrast this with the case in which we do not impose $m_{X}/H_{e}\gtrsim6$.
In that case, we can fix the value of the relic density, Eq.\
(\ref{eq:relicdensity}),
to be $\Omega_{X}h^{2}=0.1$, impose $T_{RH}<10^{9}$ GeV, and minimize
the value of $H_{e}$ to obtain $H_{e}\gtrsim3\times10^{11}\textrm{GeV}$,
where the minimzation of $H_{e}$ occurs for $m_{X}/H_{e}=5/4$.
Since $H_{e}$ is one order of magnitude smaller, the gravitational
wave signal in this case is bounded from below by a number that is
$100$ times smaller than Eq.\ (\ref{eq:grwavebound2}). Hence, the
isocurvature perturbation constraint strengthens the preference of
large tensor perturbations for the gravitationally produced
superheavy dark matter scenario.

\section{CMB temperature fluctuations and Isocurvature perturbations}

\subsection{Dynamics in the tight-coupling regime}

Here we follow the treatment of the Mukhanov \cite{Mukhanov:2003xr} (for
other good analytic treatments, see \cite{Hu:1994uz,Hu:1994jd}). The background
FRW metric is taken with flat spatial sections and conformal time,
whose time derivative is denoted with a prime ($'$). The scalar metric
perturbation parameterization is chosen to be
\begin{equation}
ds^{2}=a^{2}\left[(1+2\Phi)d\tau^{2}-(1-2\Phi)\delta_{ik}dx^{i}dx^{k}\right] .
\label{metric}
\end{equation}
Assuming that the cold dark matter (denoted with subscript $CDM$) is decoupled
from the baryon-photon plasma, the dark-matter stress-energy conservation
equations lead to
\begin{eqnarray}
(\delta_{X}-3\Phi)'+au_{CDM,i}^{i} & = & 0 \\
\label{eq:CDMcons}
[a(\delta_{X}-3\Phi)']'-a\Delta\Phi & = & 0.
\label{eq:CDMspace}
\end{eqnarray}
where $u^i_X$ is the velocity of the species $X$.\\
With the approximation of nonrelativistic baryons, the
conservation equation $T_{\,\,0;\alpha}^{\alpha}=0$ for baryons
leads to
\begin{equation}
(\delta_{b}-3\Phi)'+au_{b,i}^{i} = 0 ,
\label{eq:conserve1}
\end{equation}
which corresponds to the conservation of baryon number. The photon
energy conservation leads to
\begin{equation}
(\delta_{\gamma}-4\Phi)'+\frac{4}{3}au_{\gamma,i}^{i} = 0 .
\label{eq:conserve2}
\end{equation}
Assuming that the photons and baryons are tightly coupled, we set
$u^{i}\equiv u_{b}^{i}=u_{\gamma}^{i}$,
which from Eqs.\ (\ref{eq:conserve1}) and (\ref{eq:conserve2}) leads to
\begin{equation}
\frac{3}{4}\delta_{\gamma}-\delta_{b}=\textrm{constant}.
\end{equation}

With this tight-coupling approximation, the $T_{\,\,\, i;\alpha}^{\alpha}=0$
of the baryon-photon stress tensor leads to
\begin{equation}
\frac{1}{a^{4}}\left[a^{5}(\rho_{b}+\rho_{\gamma}+P_{\gamma}+P_{b})
u_{,i}^{i}\right]'-\frac{4}{3}\eta\Delta u_{\,\,,i}^{i}
+\Delta\delta(P_{\gamma}+P_{b})
+(\epsilon_{\gamma}+\epsilon_{b}+P_{\gamma}+P_{b})\Delta\Phi=0 ,
\end{equation}
where $\eta$ is the viscosity:
\begin{equation}
\eta=\frac{4}{15}\rho_{\gamma}\tau_{\gamma},
\end{equation}
with $\tau_{\gamma}$  the mean free path of the photons. Assuming
no isocurvature perturbations from the baryons, we use
\begin{equation}
\delta_{b}=\frac{3}{4}\delta_{\gamma}
\end{equation}
and arrive at
\begin{equation}
\left(\frac{\delta_{\gamma}'}{c_{s}^{2}}\right)'
-\frac{3\eta}{\rho_{\gamma}a}\Delta\delta_{\gamma}'
-\Delta\delta_{\gamma}=\frac{4}{3c_{s}^{2}}\Delta\Phi
+\left(\frac{4\Phi'}{c_{s}^{2}}\right)'
-\frac{12\eta}{\rho_{\gamma}a}\Delta\Phi',
\label{eq:baryonphotonspace}
\end{equation}
with the sound speed
\begin{equation}
c_{s}^{2}=\frac{1}{3}\frac{1}{1+3\rho_{b}/4\rho_\gamma}.
\end{equation}
Finally, the 00 component of the Einstein's equations is
\begin{equation}
\Delta\Phi-3\frac{a'}{a}\Phi'-3\left(\frac{a'}{a}\right)^{2}\Phi =
\frac{4\pi}{M_{pl}^{2}}a^{2}\left(\rho_{X}\delta_{X}
+\frac{\rho_{\gamma}\delta_{\gamma}}{3c_{s}^{2}}\right).
\label{eq:00einstein}
\end{equation}
The fields that need to be determined are
$\{\delta_{X},\delta_{\gamma},\Phi\}$
and the three independent equations are Eqs.\ (\ref{eq:CDMspace}),
(\ref{eq:baryonphotonspace}), and (\ref{eq:00einstein}). (The only
remaining equation simply determines $u_{CDM}^{i}$.)

As with any differential equations, these need boundary conditions.
The CDM fluctuation initial spectrum is determined by a quantum computation
of $\langle\delta_{X}\delta_{X}\rangle$ which is assumed to evolve
collisionlessly at least until decoupling. Similarly, quantum computation
of adiabatic perturbations $\langle\zeta\zeta\rangle$ provide another
set of boundary conditions. Finally, since the 00 component of Einstein's
equation, Eq.\ (\ref{eq:00einstein}), is a parabolic equation for
$\Phi$, we only require one boundary condition for $\Phi$. The fact
there is a growing solution supported by a source obviates the need
for a boundary condition for $\Phi$ for the cases of our interest.

\subsection{Relationship between CMB temperature and isocurvature fluctuation}

Let the temperature of the CMB photons be a field
$T+\Delta T(\tau,\vec{x},\hat{p})$
where $\hat{p}\equiv p_{c}^{i}/|\vec{p}_{c}|$ corresponds to
the direction of photon propagation (the subscript refers to coordinate
momentum). We would like to find $\Delta T(\tau_{0},\vec{x},\hat{p})$,
the temperature fluctuation of photons today, given $\Delta T(\tau_{i},
\vec{x},\hat{p})$,
the temperature fluctuations of photons at the time of last-scattering
surface. We can define the relativistic phase space volume as
\begin{equation}
|g_{ab}-n_{a}n_{b}|d^{3}xd^{3}p,
\end{equation}
where the absolute value signifies determinant and $n_{a}$ are timelike
vectors normal to the spacelike hypersurface. This is just a fancy
way of writing physical momentum and space volume for a fixed-time
slicing. The free-streaming Boltzmann equation for evolving $\Delta T$
is given by
\begin{equation}
p_{c}^{\mu}\partial_{\mu}f-p_{c}^{\alpha}p_{c}^{\beta}
\Gamma_{\alpha\beta}^{\mu}\partial_{p_{c}^{\mu}}f=0,
\end{equation}
with $\Gamma_{\alpha\beta}^{\mu}$ calculated using the metric of Eq.\
(\ref{metric}).
After standard manipulations, one finds the temperature field evolution
to obey
\begin{equation}
\left(\partial_{0}+\frac{p_{c}^{i}}{|\vec{p}_{c}|}\partial_{i}\right)
\left(\frac{\Delta T}{T}+\Phi\right)=2\partial_{0}\Phi.
\end{equation}
In integrating this equation for the temperature field today, the
contribution of the right-hand side of this equation is called the
integrated Sachs-Wolfe effect, and with the right-hand side neglected,
one obtains what is simply referred to as the Sachs-Wolfe effect.

If $\Delta T/T+\Phi$ is  not vanishing initially, to
leading approximation we can neglect the integrated Sachs-Wolfe effect term
$2\partial_{0}\Phi$, and the relevant Boltzmann equation is
\begin{equation}
\left(\partial_{0}+\frac{p_{c}^{i}}{|\vec{p}_{c}|}\partial_{i}\right)
\left(\frac{\Delta T}{T}+\Phi\right)=0.
\label{eq:temppotboltzmanneq}
\end{equation}
Let us consider the initial condition
\begin{equation}
\frac{\Delta T}{T}+\Phi=\left.\frac{\Delta T}{T}\right|_i
+ \left.\Phi\right|_i
\label{eq:initialcond}
\end{equation}
for this Boltzman equation where we again remind the reader that
$i$ corresponds to the last scattering surface time and $f$ corresponds
to today. Inflationary computation gives us values for $\Delta T/T+\Phi$
during radiation domination when the modes of interest had wavelengths
far outside of the horizon. We shall denote these initial conditions
with a subscript $p$ (primordial) such that for example $\Phi|_{p}$
denotes the value during radiation domination. Starting from these
primordial values, we would like to derive the initial condition Eq.\
(\ref{eq:initialcond}) in terms of the ``gauge invariant'' curvature
perturbation $\zeta$ and the isocurvature perturbation $\delta_{S}$
(see appendix for its definition) evaluated at the last scattering
surface.\footnote{The ``gauge invariant'' curvature perturbations are
usually described
by $\zeta$ or $\mathcal{R}$ which are equivalent when the scales
are far outside of the horizon. The relationship between the two are
$\zeta\equiv-\Phi+\delta\rho/3(\rho+P)=\mathcal{R}+(k/aH)^{2}
\Phi2\rho/9(\rho+P)$ and we will choose to work with $\zeta$.}
To accomplish this, we need to express everything in terms of the
set of field variables $\{\delta_{\gamma},\Phi,\delta_{X}\}$ and
then express the final result in terms of $\zeta$ and $\delta_{S}$.
To start off, the temperature variable is easy to exchange in terms
of the photon overdensity since by definition
\begin{equation}
\left.\frac{\Delta T}{T}\right|_{i}=\frac{1}{4}\left.\delta_{\gamma}\right|_i.
\label{eq:tempgammarel}
\end{equation}

From now on, we will go to the Fourier space (spatially flat
$e^{i\vec{k}\cdot\vec{x}}$
basis) and assume all of our variables now represent amplitudes in
Fourier space. We have according to $T_{\,\,\,0;\alpha}^{\alpha}=0$
(Eq.\ (\ref{eq:conserve2})) that
\begin{equation}
C_{1}\equiv\delta_{\gamma}(\tau,k)-4\Phi(\tau,k)
\label{eq:constc1}
\end{equation}
approximately a constant on large length scales.  Because the
curvature invariant $\zeta$ computed through the usual inflationary
formalism is $\left.\zeta\right|_p = - \left.\Phi\right|_p +
\left.\delta_\gamma\right|_p/4$ where the
subscript $p$ denotes the quantities are evaluated in the radiation
dominated era, we can write
\begin{equation}
\left.\zeta\right|_p =  \frac{C_1}{4}.
\label{eq:zetapeqc1}
\end{equation}
Since $\zeta$ is an adiabatic invariant, $\zeta$ will continue to
equal $C_1/4$ even after matter domination.  By
Eq.\ (\ref{eq:constc1}), we find
\begin{equation}
\zeta =  \frac{\delta_\gamma}{4} - \Phi
\label{eq:foralltime}
\end{equation}
for all time in the long-wavelength limit.

Inserting Eqs.\ (\ref{eq:tempgammarel}) and (\ref{eq:foralltime}) into
Eq.\ (\ref{eq:initialcond}), we arrive at
\begin{equation}
\left.\frac{\Delta T}{T}\right|_i+\left.\Phi\right|_i =
\frac{1}{4}\left.\delta_{\gamma}\right|_i+\left.\Phi\right|_{i} =
\zeta+2\left.\Phi\right|_{i}. \label{eq:deltzetaphi}
\end{equation}
Note that we
have here assumed that the wavelengths of interest are sufficiently
long such that radiation era quantities are valid at the last
scattering surface which is assumed to be after matter domination.

Now, we want to reexpress $\Phi$ this in terms of the isocurvature
perturbations.  As discussed in the appendix, the isocurvature perturbation
of interest is
\begin{equation}
\delta_{S}\equiv\delta_{X}-\frac{3}{4}\delta_{\gamma}.
\end{equation}
Hence, we need to eliminate $\delta_X$ and
$\delta_\gamma$ in terms of $\delta_s$ and $\zeta$.

To relate $\delta_X$ with $\Phi$, we can use the energy
conservation equation (Eq.\ (\ref{eq:CDMcons})) in the
long-wavelength limit, and restricting to the non decaying mode,
to write
\begin{equation}
\delta_{X}-3\Phi =C_2,
\label{eq:constx}
\end{equation}
where $C_2$ is a constant.  Now, since during matter domination, the
00 component of Einstein's equation (Eq.\ (\ref{eq:00einstein})) in the
long-wavelength limit is
\begin{equation}
\left.\Phi'\right|_{i}+\frac{a'}{a}\left.\Phi\right|_{i}
=-\frac{1}{2}\frac{a'}{a}\left.\delta_{X}\right|_{i} ,
\end{equation}
we can use Eq.\ (\ref{eq:constx}) to solve for the growing solution,
which is
\begin{equation}
\left.\delta_{X}\right|_{i}\approx-2\left.\Phi\right|_{i}.
\label{eq:duringmd}
\end{equation}
During matter domination, we thus find using Eqs.\ (\ref{eq:duringmd})
and (\ref{eq:foralltime}) that
\begin{equation}
\delta_{S} =
-2\left.\Phi\right|_{i}-3\left(\zeta+\left.\Phi\right|_{i}\right) =
-5\left.\Phi\right|_{i}-3\zeta.
\label{eq:isocurvaturesourcingpot}
\end{equation}
This is a remarkable feature of the isocurvature perturbations in
which even if $\zeta=0$, the isocurvature perturbations grow into
potential perturbations during matter domination.

We say that the isocurvature perturbations grow into potential
perturbations because during radiation domination, if $\zeta=0$, then
the gravitational perturbations approximately vanish.  Let us see how
this happens in detail.  Solving the 00 part of the Einstein equation
(Eq.\ (\ref{eq:00einstein})), we find during radiation domination (RD)
that
\begin{equation}
\left.\Phi\right|_{p}(\tau,k)=\frac{-C_{1}}{6} ,
\label{eq:potduringraddom}
\end{equation}
where as before, the subscript $p$ denotes the radiation
domination era when the wavelengths of interest are far outside
the horizon. This and Eq.\ (\ref{eq:zetapeqc1}) imply that in the
absence of curvature perturbations $\zeta$, the gravitational
potential $\Phi$ is 0 during radiation domination. However, during
matter domination, the potential $\Phi$ grows due to the existence
of the isocurvature perturbations. One can also understand from
this (the fact that during radiation domination
$\Phi=\delta_\gamma=0$ if $\zeta=0$) that if matter domination
never occurs then we would have in a hypothetical radiation
dominated last scattering surface that
\begin{equation}
\left.\frac{\Delta T}{T}\right|_{i \mbox{ hypothetical}}
+ \left.\Phi\right|_{i \mbox{ hypothetical}} =0
\end{equation}
which is in accord with the intuition that the superheavy dark matter
perturbations are irrelevant if their density is too small.

Combining Eq.\ (\ref{eq:isocurvaturesourcingpot}) with Eq.\
(\ref{eq:deltzetaphi}),
we arrive at the desired expression
\begin{equation}
\left.\frac{\Delta T}{T}\right|_{i}+\left.\Phi\right|_{i}= \frac{-1}{5}\zeta
-\frac{2}{5}\delta_{S}.
\label{eq:temperatureatlastscattering}
\end{equation}

\subsection{$C_{l}$ characterization of CMB}

Now consider the computation of $C_{l}$ which is defined as
\begin{equation}
\langle a_{lm}a_{l'm'}^{*}\rangle=\delta_{ll'}\delta_{mm'}C_{l} ,
\end{equation}
where
\begin{equation}
\frac{\Delta T(\tau_{f},\vec{x},\hat{p})}{T} =
\sum_{l,m}a_{lm}(\tau_{f},\vec{x})Y_{lm}(\hat{p})
\end{equation}
and $\hat{p}$ is the direction vector of the photon. Using
\begin{equation}
a_{lm}(\vec{x},\tau_{f})=
\int\frac{d^3k}{(2\pi)^{3}}e^{i\vec{k}\cdot\vec{x}}\int
d\Omega \ Y_{lm}^{*}(\hat{p})\frac{\Delta T(\tau_{f},\vec{x},\hat{p})}{T} ,
\end{equation}
where $\Omega$ is the solid angle for $\hat{p}$, we write the ensemble
average for $C_{l}$ as
\begin{eqnarray}
\langle a_{lm}a_{l'm'}^{*}\rangle & = & \int\frac{d^{3}k_{1}d^{3}k_{2}}
{(2\pi)^{6}}\int d\Omega_{1}\int d\Omega_{2}\left\langle
\frac{\Delta T(\tau_{f},
\vec{k}_{1},\hat{p}_{1})}{T}\frac{\Delta T^{*}(\tau_{f},\vec{k}_{2},
\hat{p}_{2})}{T}\right\rangle e^{i\vec{k}_{1}\cdot\vec{x}}e^{-i\vec{k}_{2}
\cdot\vec{x}}Y_{lm}^{*}(\hat{p}_{1})Y_{l'm'}(\hat{p}_{2}) ,
\end{eqnarray}
where we have gone to spatial Fourier space. Now, use the property
that after the last scattering surface the Boltzmann equation Eq.\
(\ref{eq:temppotboltzmanneq}) for long wavelengths is approximately
\begin{equation}
\left(\partial_{0}+\hat{p}^{i}\partial_{i}\right)\left(\frac{\Delta
T}{T}+\Phi\right) \approx 0
\end{equation}
to write
\begin{equation}
\frac{\Delta T(\tau,\vec{x},\hat{p})}{T}+\Phi(\tau,\vec{x})
\approx \int\frac{d^{3}k}{(2\pi)^{3}} \left(\left.\frac{\Delta
T(\vec{k})}{T}\right|_{i}+\ \left.\Phi(\vec{k})\right|_{i}\right)
e^{i\vec{k}\cdot\vec{x}} e^{-i\hat{p}\cdot\vec{k}(\tau-\tau_{i})}.
\end{equation}
Neglecting $\Phi(\tau_{f},\vec{x})$ (today) gives
\begin{equation}
\frac{\Delta T(\tau_{f},\vec{k},\hat{p}_{1})}{T} \approx
\left( \left.\frac{\Delta T(\vec{k})}{T}\right|_{i}+\Phi(\vec{k})|_{i}
\right) e^{-i\hat{p}_{1}\cdot\vec{k}(\tau_{f}-\tau_{i})}.
\end{equation}
Hence, we find
\begin{eqnarray}
\langle a_{lm}a_{l'm'}^{*}\rangle & = & \frac{1}{V}\int\frac{d^{3}k_{1}}
{(2\pi)^{3}} \frac{d^{3}k_{2}}{(2\pi)^{3}}\int d\Omega_{1}\int d\Omega_{2}
\left\langle\left(\left.\frac{\Delta T(\vec{k}_{1})}{T}\right|_{i} +
\left.\Phi(\vec{k}_{1})\right|_{i}\right)\left(\left.
\frac{\Delta T^{*}(\vec{k})}{T}\right|_{i} +
\left.\Phi^{*}(\vec{k}_{2})\right|_{i}\right)\right\rangle \\ \nonumber
 &  & \times e^{-i\hat{p}_{1}\cdot\vec{k}_{1}(\tau_{f}-\tau_{i})}
 e^{i\hat{p}_{2}\cdot\vec{k}_{2}(\tau_{f}-\tau_{i})}e^{i\vec{k}_{1}
 \cdot\vec{x}}e^{-i\vec{k}_{2}\cdot\vec{x}}Y_{lm}^{*}(\hat{p}_{1})
 Y_{l'm'}(\hat{p}_{2}).
\end{eqnarray}

To leading order, the power temperature correlation function can be
written as
\begin{equation}
\left\langle\left(\left.\frac{\Delta T(\vec{k}_{1})}{T}\right|_{i}
+\left.\Phi(\vec{k}_{1})\right|_{i}\right)
\left(\left.\frac{\Delta T^{*}(\vec{k}_{2})}{T}\right|_{i}
+\left. \Phi^{*}(\vec{k}_{2})\right|_{i}\right) \right\rangle
=\frac{2\pi^{2}}{k_{1}^{3}}P(k_{1})(2\pi)^{3}\delta^{(3)}
(\vec{k}_{1}-\vec{k}_{2}).\label{eq:defofpower}
\end{equation}
Inserting
\begin{equation}
e^{-i\hat{p}_{1}\cdot\vec{k}(\tau_{f}-\tau_{i})}=
\sum_{l=0}^{\infty}i^{l}(2l+1)j_{l}(|\vec{k}(\tau_{f}-\tau_{i})|)
P_{l}(-\hat{p}_{1}\cdot\hat{k})
\end{equation}
where $P_{l}$ are Legender polynomials, we find
\begin{eqnarray}
\langle a_{lm}a_{l'm'}^{*}\rangle & = &
\int\frac{d^{3}k}{(2\pi)^{3}} \int d\Omega_{1}\int
d\Omega_{2}\frac{2\pi^{2}}{k^{3}}P(k)
\sum_{l_{1}=0}^{\infty}i^{l_{1}}(2l_{1}+1)j_{l_{1}}(|\vec{k}(\tau_{f}-
\tau_{i})|)P_{l_{1}}(-\hat{p}_{1}\cdot\hat{k})Y_{lm}^{*}(\hat{p}_{1}) \\
\nonumber
&  & \times \sum_{l_{2}=0}^{\infty}i^{l_{2}}(2l_{2}+1)j_{l_{2}}
(|\vec{k}(\tau_{f}-\tau_{i})
|)P_{l_{2}}(\hat{p}_{2}\cdot\hat{k})Y_{l'm'}(\hat{p}_{2}).
\end{eqnarray}
Now using the identity
\begin{equation}
P_{l}(-\hat{p}_{1}\cdot\vec{k})=\frac{4\pi}{2l+1}
\sum_{m=-l}^{l}Y_{lm}(-\hat{p}_{1})Y_{lm}^{*}(\hat{k}) ,
\end{equation}
where $Y_{lm}$ are orthonormal spherical harmonics, we find
\begin{eqnarray}
\langle a_{lm}a_{l'm'}^{*}\rangle & = & \delta_{ll'}\delta_{mm'}4\pi
\int\frac{dk}{k}P(k)[j_{l}(|\vec{k}(\tau_{f}-\tau_{i})|)]^{2}.
\end{eqnarray}
Thus, we see that $C_{l}$ is essentially the power spectrum on long
wavelengths up to angular projection effects of the Bessel function.
Defining the photon travel distance from the last scattering surface
\begin{equation}
L \equiv \tau_{0}-\tau_{dec}
\approx \frac{1}{H_{0}a_{0}}\int_{0}^{z_{dec}}
\frac{dz}{\sqrt{\Omega_{m}(z+1)^{3}+\Omega_{\Lambda}}} ,
\label{eq:defofdisttols}
\end{equation}
we arrive at the desired expression for $C_{l}$ as
\begin{equation}
C_{l}=4\pi\int\frac{dk}{k}P(k)[j_{l}(kL)]^{2}.
\label{eq:finalcl}
\end{equation}
To evaluate $C_{l}$ from this expression, we merely need to compute
$P(k)$, which from Eqs.\ (\ref{eq:defofpower}) and
(\ref{eq:temperatureatlastscattering})
is seen to be
\begin{equation}
P(k)=\frac{1}{25}\left.P_{\zeta}(k)\right|_{i}
+\frac{4}{25}\left.P_{S}(k)\right|_{i} ,
\label{eq:totalpowerspectrumform}
\end{equation}
where the subscript $i$ denotes last scattering surface.

To compute $P_S$ at the last scattering surface, from
Eqs.\ (\ref{eq:constc1}) and (\ref{eq:constx}) we find
\begin{equation}
\delta_S =C_2 - \frac{3}{4} C_1.
\label{eq:constancyofs}
\end{equation}
In other words, isocurvature perturbations are approximately constant
far outside of the horizon, whether the quantities are evaluated in
matter or radiation domination era.  Therefore, since we can express
$\delta_\gamma$ in terms of $\zeta_p$ in radiation domination using
Eqs.\ (\ref{eq:constc1}), (\ref{eq:zetapeqc1}), and
(\ref{eq:potduringraddom}) as
\begin{equation}
\zeta_p = \frac{3}{4} \delta_{\gamma p} ,
\label{eq:zetadelgammaprelation}
\end{equation}
it is convenient to compute $P_S$ in the radiation domination era
using the relationship
\begin{equation}
\left.\delta_S\right|_p = \left.\delta_X\right|_p - \left.\zeta\right|_p ,
\label{eq:sdeltaandzetarelated}
\end{equation}
where one should understand the existence of nonzero $\delta_S|_p$
even when $\delta_X|_p=0$ from the fact that the dark matter number
density would then not trace the radiation number density.  Hence,
$P_{\zeta}(k)$ and $P_{S}(k)$ are
\begin{eqnarray}
\left.P_S(k)\right|_i & = & \left.P_{\delta_{X}}\right|_p +
\left.P_{\zeta}\right|_p - \left.P_{\zeta X}\right|_p
- \left.P_{X\zeta}\right|_p  \\
P_{\delta_{X}} & = & \frac{k^{3}}{2\pi^{2}}\int
d^{3}r\ e^{-i\vec{k}\cdot\vec{r}}\langle\delta_{X}(\vec{x})\delta_{X}(\vec{y})
\rangle
\label{eq:delxpower}
\\
P_{\zeta}(k) & = &\frac{k^{3}}{2\pi^{2}}\int d^{3}r \ e^{-i\vec{k}\cdot\vec{r}}
\langle\zeta(\vec{x})\zeta(\vec{y})\rangle \\
P_{\zeta X}(k) & = & \frac{k^{3}}{2\pi^{2}}
\int d^{3}r \ e^{-i\vec{k}\cdot\vec{r}}\langle\zeta(\vec{x})\delta_X(\vec{y})
\rangle ,
\end{eqnarray}
where all the functions with subscript $p$ are evaluated at the
radiation-dominated era.  For the usual thermal CDM scenario, we have
\begin{equation}
\delta_X= \frac{3}{4} \delta_\gamma,
\end{equation}
which gives a zero power for the isocurvature as expected.  For
the isocurvature of superheavy dark matter, the cross correlation
terms $P_{\zeta X}$ and $P_{X \zeta}$ vanish.  The first term of
Eq.\ (\ref{eq:totalpowerspectrumform}) is the usual adiabatic
contribution to the CMB which can be easily computed for slow roll
inflationary scenarios in a standard way. The second term of Eq.\
(\ref{eq:totalpowerspectrumform}) (or more specifically Eq.\
(\ref{eq:delxpower})) is what we are primarily concerned with in
this paper, and we will turn to its computation in the next
section.

Before beginning the computation of the matter correlation function,
note that Eq.\ (\ref{eq:zetadelgammaprelation}) allows us to rewrite
the familiar Eq.\ (\ref{eq:totalpowerspectrumform}) as
\begin{equation}
P(k)= \frac{1}{5} P_{\zeta } +\frac{4}{25} P_{\delta_X } ,
\label{eq:bettertotpowerspecform}
\end{equation}
where we have dropped the subscript for brevity.

\section{Isocurvature power spectrum}

In this section, we proceed to compute $P_{\delta_X}(k)$ needed in
Eq.\ (\ref{eq:bettertotpowerspecform}) for the computation of $C_{l}$
using Eq.\ (\ref{eq:finalcl}).  As we discussed in the introduction,
we restrict ourselves to a massive scalar field $X$ minimally coupled
to gravity without any other interactions for $X$.  The quantum fluctuations
of the energy density during inflation is usually computed by using
the curvature perturbation which remains constant far outside the
horizon. However, the usual computation procedure which attempts to
compute the quantum fluctuations induced about a \emph{classical field
background} does not apply to our scenario since the dark matter
particles do not arise from a classical field background but from
quantum particle production. Liddle and Mazumdar \cite{Liddle:1999pr}
had a similar cosmological scenario as the one of interest in this
paper, but instead of real particle production, their homogeneous dark
matter density was implicitly from the vacuum energy contribution.
This difference can be readily seen by computing the vacuum
expectation value of the stress energy tensor for a massive scalar
field $X$ with mass $m_{X}$ as
\begin{equation}
\int d^{3}x \ \langle T_{00}(x)\rangle =
\int d^{3}k \ a^{3}\langle A_{\vec{k}}A_{\vec{k}}^{\dagger}\rangle
\left[\left(|\beta_{\vec{k}}|^{2}+\frac{1}{2}\right)
\left(|\dot{X}_{k}(t)|^{2}+(\vec{k}^{2}/a^{2}+m_{X}^{2})|X_{k}(t)|^{2}\right)
\right],
\end{equation}
where the vacuum state $|0\rangle$ is defined such that
$A_{\vec{k}}|0\rangle=0$
while the fields are defined by a \emph{Bogoliubov rotated} boundary
condition. The vacuum of Ref.\ \cite{Liddle:1999pr} corresponds
to the case with $\beta_{k}=0$, and their energy density came from
the $\frac{1}{2}m^{2}|X_{k}|^{2}$ term in the integral, which is
part of the zero point energy term.

Given that the zero point energy must be taken care of by the cosmological
constant problem solution, whether the zero point energy should be
counted as an unambiguous production of particles is unclear. Indeed,
in Minkowski space, we usually discard the zero point energy. One
might argue that the zero point energy is what is contributing to
the usual computation of generating density fluctuations in inflationary
cosmology. However, this is not true. As we stated previously, the
fluctuations that are being computed in the usual density perturbation
computations are not the fluctuations about the classically zero energy
state but about a quasi-de Sitter background of positive energy. Indeed,
in computing the density fluctuations, the zero point energy of the
inflaton is always subtracted. Hence, since treating the zero point
energy as real particle production without specifying assumptions
about the solution to the cosmological constant problem is speculative,
we will in our computations throw away the zero point energy just
as in Minkowski space. This is one of several important differences
between Liddle and Mazumdar \cite{Liddle:1999pr} and our paper.

Assuming that the inflaton energy density dominates over the dark
matter energy density during inflation, the curvature perturbations
will be computed in the usual manner.  This is an excellent assumption
since the dark matter energy density to inflaton energy density during
inflation needs to be much smaller than $10^{-10}$ to obtain the
phenomenologically acceptable amount of dark matter
today. Furthermore, because the superheavy dark matter field is
assumed to be sitting at the minimum of its potential with zero vacuum
energy, there is no vacuum energy contribution coming from the
superheavy dark matter sector unlike the inflaton sector.  In other
words, the energy perturbation $\delta \rho$ coming from inflation is
linear in the inflaton field fluctuation $\delta \phi$ while the
energy perturbation coming from the superheavy dark matter is
quadratic in $X$.  Hence, to linear order, there is no mixing between
the inflaton energy density and the superheavy dark matter energy
density.

The dark matter energy density fluctuation is defined as
\begin{equation}
\delta \rho_X (\vec{x}) = : T_{00}^{(X)}(\vec{x}) : - \langle :
  T_{00}^{(X)} : \rangle,
\label{eq:energydensitydef}
\end{equation}
where the normal ordering is with respect to the Bogoliubov rotated
operators to eliminate the vacuum energy.  Hence, we can write the
correleation function as
\begin{equation}
\langle\delta\rho_{X}(\vec{x})\delta\rho_{X}(\vec{y})\rangle
=\langle:T_{00}^{(X)}(\vec{x}): \ :T_{00}^{(X)}(\vec{y}):\rangle
- \langle : T_{00}^{(X)} :\rangle^2.
\label{eq:twopointdelrhodef}
\end{equation}
Explicitly, the Bogoliubov rotation is given by
\begin{equation}
a_{\vec{k}}=\alpha_{\vec{k}}A_{\vec{k}}
+\beta_{-\vec{k}}^{*}A_{-\vec{k}}^{\dagger} ,
\label{eq:bogo}
\end{equation}
and the normal ordered stress tensor vacuum expectation value is given
as
\begin{equation}
\rho_X(t) \equiv \langle:T_{00}^{(X)}:\rangle  =
\int\frac{d^{3}k_{1}}{(2\pi)^{3}}
\left\{\Re\left[\alpha_{\vec{k_{1}}}\beta_{\vec{k}_{1}}^{*}
\left(\dot{X}_{k_{1}}^{2}(t) +w_{k_{1}}^{2}X_{k_{1}}^{2}(t)\right)\right]
+|\beta_{\vec{k}_{1}}|^{2}\left(|\dot{X}_{k_{1}}(t)|^{2}
+w_{k_{1}}^{2}|X_{k_{1}}(t)|^{2}\right)\right\} ,
\end{equation}
where $[dk]=d^{3}k/(2\pi)^{3/2}$ and we have used the fact
that $X_{\vec{k}_{1}}=X_{-\vec{k}_{1}}$ and defined
$w_{k}^{2}=\vec{k}^{2}/a^{2}+m^{2}$.  Note that this vanishes
in the limit $\beta_{\vec{k}}\rightarrow0$ where there is no particle
production consistent with our previous discussion of comparing our
paper with \cite{Liddle:1999pr}.
Hence, the dark matter correlation function is
\begin{equation}
\langle \delta_X(\vec{x}) \delta_X(\vec{y}) \rangle =
\frac{\langle:T_{00}^{(X)}(\vec{x}): \ :T_{00}^{(X)}(\vec{y}):
  \rangle}{\rho_X^2} -1. \label{eq:corrwithmin1}
\end{equation}

With the mode mixing given by Eq.\ (\ref{eq:bogo}), the stress tensor
correlator can be written down straightforwardly
\begin{eqnarray}
& & \hspace{-24pt}
\left\langle:T_{00}^{(X)}(\vec{x}): \ :T_{00}^{(X)}(\vec{y}):\right\rangle =
\nonumber \\ & & \ \ \hspace{-12pt}
\left\langle \frac{1}{2} \int[dk_{1}][dk_{2}]
\left[ \phantom{\frac{k_{1_{i}}k_{2_{i}}}{a^{2}}} \hspace*{-30pt}
\left( \alpha_{\vec{k}_{1}} \alpha_{\vec{k}_{2}}A_{\vec{k}_{1}}
A_{\vec{k}_{2}} + \beta_{-\vec{k}_{1}}^{*}\alpha_{\vec{k}_{2}}
A_{-\vec{k}_{1}}^{\dagger}A_{\vec{k}_{2}}
+ \alpha_{\vec{k}_{1}}\beta_{-\vec{k}_{2}}^{*}A_{\vec{k}_{1}}
A_{-\vec{k}_{2}}^{\dagger}
+ \beta_{-\vec{k}_{1}}^{*}\beta_{-\vec{k}_{2}}^{*}
A_{-\vec{k}_{1}}^{\dagger}A_{-\vec{k}_{2}}^{\dagger}\right) \right. \right.
\nonumber \\ &  & \hspace{72pt}
\times \left( \dot{X}_{k_{1}}(t)\dot{X}_{k_{2}}(t)
-\frac{k_{1_{i}}k_{2_{i}}}{a^{2}}X_{k_{1}}(t)X_{k_{2}}(t)
+m^{2}X_{k_{1}}(t)X_{k_{2}}(t)\right)e^{i(\vec{k}_{1}+\vec{k}_{2})\cdot\vec{x}}
\nonumber \\ & & \hspace{56pt}
+ 2 \left(\alpha_{\vec{k}_{1}}^{*}\alpha_{\vec{k}_{2}}A_{\vec{k}_{1}}^{\dagger}
A_{\vec{k}_{2}}+\beta_{-\vec{k}_{1}}\alpha_{\vec{k}_{2}}A_{-\vec{k}_{1}}
A_{\vec{k}_{2}}+\alpha_{\vec{k}_{1}}^{*}\beta_{-\vec{k}_{2}}^{*}
A_{\vec{k}_{1}}^{\dagger}A_{-\vec{k}_{2}}^{\dagger}+\beta_{-\vec{k}_{1}}
\beta_{-\vec{k}_{2}}^{*}A_{-\vec{k}_{1}}A_{-\vec{k}_{2}}^{\dagger}\right)
\nonumber \\ & &  \hspace{72pt}
\times \left( \dot{X}_{k_{1}}^{*}(t)\dot{X}_{k_{2}}(t)
+\frac{k_{1_{i}}k_{2_{i}}}{a^{2}} X_{k_{1}}^{*}(t)X_{k_{2}}(t)
+m^{2}X_{k_{1}}^{*}(t)X_{k_{2}}(t) \right)
e^{-i(\vec{k}_{1}-\vec{k}_{2})\cdot\vec{x}}
\nonumber \\ & & \hspace{56pt}
+ \left(\alpha_{\vec{k}_{1}}^{*}\alpha_{\vec{k}_{2}}^{*}
A_{\vec{k}_{1}}^{\dagger}
A_{\vec{k}_{2}}^{\dagger}+\beta_{-\vec{k}_{1}}\alpha_{\vec{k}_{2}}^{*}
A_{-\vec{k}_{1}}A_{\vec{k}_{2}}^{\dagger}+\alpha_{\vec{k}_{1}}^{*}
\beta_{-\vec{k}_{2}}A_{\vec{k}_{1}}^{\dagger}A_{-\vec{k}_{2}}+
\beta_{-\vec{k}_{1}}\beta_{-\vec{k}_{2}}A_{-\vec{k}_{1}}A_{-\vec{k}_{2}} \right)
\nonumber \\ & & \hspace{72pt} \left.
\times \left( \dot{X}_{k_{1}}^{*}(t)\dot{X}_{k_{2}}^{*}(t)
-\frac{k_{1_{i}}k_{2_{i}}}{a^{2}}X_{k_{1}}^{*}(t)X_{k_{2}}^{*}(t)
+m^{2}X_{k_{1}}^{*}(t)X_{k_{2}}^{*}(t)\right)
e^{-i(\vec{k}_{1}+\vec{k}_{2})\cdot\vec{x}} \right]
\nonumber  \\ & &  \ \ \ \  \hspace{-12pt}
\frac{1}{2}\int[dk_{3}][dk_{4}]
\left[\phantom{\frac{k_{1_{i}}k_{2_{i}}}{a^{2}}} \hspace*{-30pt}
\left(\alpha_{\vec{k}_{3}}\alpha_{\vec{k}_{4}}
A_{\vec{k}_{3}}A_{\vec{k}_{4}}+\beta_{-\vec{k}_{3}}^{*}\alpha_{\vec{k}_{4}}
A_{-\vec{k}_{3}}^{\dagger}A_{\vec{k}_{4}}+\alpha_{\vec{k}_{3}}
\beta_{-\vec{k}_{4}}^{*}A_{\vec{k}_{3}}A_{-\vec{k}_{4}}^{\dagger}+
\beta_{-\vec{k}_{3}}^{*}\beta_{-\vec{k}_{4}}^{*}A_{-\vec{k}_{3}}^{\dagger}
A_{-\vec{k}_{4}}^{\dagger}\right) \right.
\nonumber  \\ & &  \hspace{72pt}
\times \left(\dot{X}_{k_{3}}(t)\dot{X}_{k_{4}}(t)-\frac{k_{3_{i}}k_{4_{i}}}
{a^{2}} X_{k_{3}}(t)X_{k_{4}}(t)+m^{2}X_{k_{3}}(t)X_{k_{4}}(t)\right)
e^{i(\vec{k}_{3}+\vec{k}_{4})\cdot y}
\nonumber \\ & &  \hspace{56pt}
+2\left(\alpha_{\vec{k}_{3}}^{*}\alpha_{\vec{k}_{4}}A_{\vec{k}_{3}}^{\dagger}
A_{\vec{k}_{4}}+\beta_{-\vec{k_{3}}}\alpha_{\vec{k}_{4}}A_{-\vec{k}_{3}}
A_{\vec{k}_{4}}+\alpha_{\vec{k}_{3}}^{*}\beta_{-\vec{k}_{4}}^{*}
A_{\vec{k}_{3}}^{\dagger}A_{-\vec{k}_{4}}^{\dagger}+\beta_{-\vec{k}_{3}}
\beta_{-\vec{k}_{4}}^{*}A_{-\vec{k}_{3}}A_{-\vec{k}_{4}}^{\dagger}\right)
\nonumber  \\ & & \hspace{72pt}
\times \left(\dot{X}_{k_{3}}^{*}(t)\dot{X}_{k_{4}}(t)
+\frac{k_{3_{i}}k_{4_{i}}}{a^{2}}
X_{k_{3}}^{*}(t)X_{k_{4}}(t)+m^{2}X_{k_{3}}^{*}(t)X_{k_{4}}(t)\right)
e^{-i(\vec{k}_{3}-\vec{k}_{4})\cdot\vec{y}}
\nonumber  \\ & & \hspace{56pt}
+ \left(\alpha_{\vec{k}_{3}}^{*}\alpha_{\vec{k}_{4}}^{*}
A_{\vec{k}_{3}}^{\dagger}
A_{\vec{k}_{4}}^{\dagger}+\beta_{-\vec{k}_{3}}\alpha_{\vec{k}_{4}}^{*}
A_{-\vec{k}_{3}}A_{\vec{k}_{4}}^{\dagger}+\alpha_{\vec{k}_{3}}^{*}
\beta_{-\vec{k}_{4}}A_{\vec{k}_{3}}^{\dagger}A_{-\vec{k}_{4}}+
\beta_{-\vec{k}_{3}}\beta_{-\vec{k}_{4}}A_{-\vec{k}_{3}}A_{-\vec{k}_{4}}\right)
\nonumber \\ & & \hspace{72pt}  \left. \left.
\times \left(\dot{X}_{k_{3}}^{*}(t)\dot{X}_{k_{4}}^{*}(t)-
\frac{k_{3_{i}}k_{4_{i}}}{a^{2}}X_{k_{3}}^{*}(t)X_{k_{4}}^{*}(t)+
m^{2}X_{k_{3}}^{*}(t)X_{k_{4}}^{*}(t)\right)
e^{-i(\vec{k}_{3}+\vec{k}_{4})\cdot y} \right] \right\rangle
\end{eqnarray}
Although not rigorously justified, we can
approximate to within an order of magnitude $|\beta_{k}|\ll
|\alpha_k|$. In the expansion with $\beta_{k}\rightarrow0$, the
correlator is simply
\begin{eqnarray}
\left\langle:T_{00}^{(X)}(\vec{x}): \ :T_{00}^{(X)}(\vec{y}): \right\rangle
& = & \frac{1}{2} \int\frac{d^{3}k_{1}d^{3}k_{2}}{(2\pi)^{6}}|
\alpha_{k_{1}}|^{2}|
\alpha_{k_{2}}|^{2}\left|\dot{X}_{k_{1}}\dot{X}_{k_{2}}-\frac{\vec{k}_{1}
\cdot\vec{k}_{2}}{a^{2}}X_{k_{1}}X_{k_{2}}+m^{2}X_{k_{1}}X_{k_{2}}
\right|^{2}e^{i(\vec{k}_{1}+\vec{k}_{2})\cdot(\vec{x}-\vec{y})}
\nonumber \\ & & +O(\beta_{k}).
\end{eqnarray}
The power spectrum is
\begin{eqnarray} P_{\delta_{X}}(k) & = &
\frac{k^{3}}{2\pi^{2}}\int
d^{3}re^{-i\vec{k}\cdot\vec{r}}\langle\delta_X(x)\delta_X(y)\rangle
= \frac{1}{\rho_{X}^{2}}\frac{k^{3}}{2\pi}\frac{1}{(2\pi)^{3}}\int
dk_{1}k_{1}^{2}\int_{-1}^{1}d\cos\theta
\left|\dot{X}_{k_{1}}\dot{X}_{\sqrt{k^{2}+k_{1}^{2}-2k_{1}k\cos\theta}}
\right. \nonumber \\ & & \left.
+a^{-2}\left(|\vec{k}_{1}|^{2}-|\vec{k}_{1}||\vec{k}|\cos\theta\right)
X_{k_{1}}X_{\sqrt{k^{2}+k_{1}^{2}-2k_{1}k\cos\theta}}+m^{2}X_{k_{1}}
X_{\sqrt{k^{2}+k_{1}^{2}-2k_{1}k\cos\theta}}\right|^{2}.
\label{eq:leadingpower}
\end{eqnarray}
Note that the negative unity in Eq.\ (\ref{eq:corrwithmin1}) disappears
because
\begin{equation}
-\frac{k^3}{2 \pi^2} \int d^3r e^{-i \vec{k}\cdot\vec{r}} =0.
\end{equation}

In the flat space limit, we find
\begin{equation}
\left\langle:T_{00}^{(X)}(\vec{x}): \ :T_{00}^{(X)}(\vec{x}):
\right\rangle
=\frac{1}{2}\int\frac{d^{3}k_{1}d^{3}k_{2}}{(2\pi)^{6}}
\frac{1}{(2E_{k_{1}})(2E_{k_{2}})}\left|-E_{k_{1}}E_{k_{2}}
-\frac{\vec{k}_{1}\cdot\vec{k}_{2}}{a^{2}}+m^{2}\right|^{2}.
\end{equation}
If we neglect the kinetic terms, we find
\begin{equation}
\left\langle:T_{00}^{(X)}(\vec{x})::T_{00}^{(X)}(\vec{x}): \right\rangle =
\frac{m^{4}}{2}\int\frac{d^{3}k_{1}d^{3}k_{2}}{(2\pi)^{6}}
\frac{1}{(2E_{k_{1}})(2E_{k_{2}})}= \frac{m^{4}}{2}\langle X(x)X(x)\rangle^{2},
\end{equation}
which is what we expect since in that limit $\rho\sim m^{2}X^{2}$
and
\begin{equation}
\langle:X^{2}(x): \ :X^{2}(y):\rangle=2\langle X(x)X(y)\rangle^{2}.
\end{equation}
 Note that in general, the kinetic part of the correlation function
cannot be neglected.

To evaluate the correlation function in the curved spacetime of
interest, we must compute the wave function, which solves the
following Klein Gordon differential equation:
\begin{equation}
X_k''(\tau)+2H X_k'(\tau)+\left(k^2+a^2(\tau) m\right)X_k(\tau)=0
\end{equation}
Note that, since the Dark Matter field has no VEV, the equation of
motion does not couple to the metric
fluctuations at first order.\\
We write the evolution of the scale factor in conformal time
$\tau$ as
\begin{equation}
\frac{1}{a}\frac{d}{d\tau}H=-\epsilon H^{2},
\end{equation}
where $H=a^{-2}da/d\tau$, and $\epsilon$ is the usual slow roll
parameter (potential derivatives) which we take to be
approximately constant (This is true at first order in the slow
roll parameters). With the boundary condition that
$a(\tau_{i})=a_{i}=-1/(H_{I}\tau_{i})$, we find
\begin{equation}
a=-\frac{1}{H_{I}\tau}\left\{1+\epsilon\left[
\left(1-\frac{\tau_{i}}{\tau}\right)
-\ln\frac{\tau}{\tau_{i}}\right]\right\} ,
\end{equation}
where one recognizes the usual dS scale factor $-1/(H_{I}\tau)$ in
the limit that $\epsilon\rightarrow0$. However, using this would
make an analytic solution impossible since $a^{2}$ will have
$\frac{1}{\tau^{3}}$
terms as well as $\ln(\tau/\tau_{i})$ terms while $a''/a$ will have
$1/\tau^{3}$ terms. Hence, we will use for the dark matter correlation
function in the dS approximation.

Still, we can approximately take into account the changing $H$ for
the isocurvature perturbations during slow roll inflation as follows.
Since the amplitude of the perturbations are approximately frozen
when the physical wavelength crosses the horizon, we will assume that
$H$ relevant for the density perturbation to be $k$ dependent such
that
\begin{equation}
\frac{k}{a(t_{k})H_{k}}=1\label{eq:defofkdeph} ,
\end{equation}
where $t_{k}$ corresponds to the time at which $k/a=H_{k}.$ Now,
solving the slow roll equation
\begin{equation}
\frac{\dot{H}}{H^{2}}=-\epsilon,
\end{equation}
we find
\begin{equation}
-\frac{1}{H_{k}}+\frac{1}{H_{I}}=-\epsilon(t_{k}-t_{I}),
\end{equation}
where $H_{I}$ corresponds to the expansion rate at some initial time
$t_{I}$. Using approximate dS expansion
\begin{equation}
(t_{k}-t_{I})=\frac{1}{H_{I}}\ln\left(\frac{a_{k}}{a_{I}}\right),
\end{equation}
we write
\begin{equation}
-\frac{1}{H_{k}}+\frac{1}{H_{I}}=-\frac{\epsilon}{H_{I}}
\ln\left(\frac{a_{k}}{a_{I}}\right),
\end{equation}
or equivalently
\begin{equation}
H_{k} \approx H_{I}\left(\frac{a_{k}}{a_{I}}\right)^{-\epsilon}
\end{equation}
Using Eq.\ (\ref{eq:defofkdeph}), we arrive at
\begin{equation}
H_{k} = H_{I}\left(\frac{k}{a_{I}H_{I}}\right)^{-\epsilon}+O(\epsilon^{2}) .
\label{eq:hkasfuncofhi}
\end{equation}

Hence, the mode function can be approximated as
\begin{equation}
X_{k}=\frac{\sqrt{\pi}}{2a^{3/2}\sqrt{H_{k}}}
e^{i\frac{\pi}{2}(\nu+1/2)}H_{\nu}^{(1)}\left(\frac{k}{aH_{k}}\right) ,
\end{equation}
where
\begin{equation}
\nu \equiv \sqrt{\frac{9}{4}-\left(\frac{m_{X}}{H_k}\right)^{2}}.
\label{eq:nufirst}
\end{equation}
The mode functions satisfy the usual normalization
\begin{equation}
2\Im\left(\dot{X}_{k}^{*}X_{k}\right)=a^{-3}.
\end{equation}
In the long-wavelength limit, we find
\begin{equation}
X_{k} \approx -\frac{1}{\sqrt{\pi}}(-1)^{3/4}2^{-1-\nu}e^{-i\nu\pi/2}
e^{-H_{k}t(3/2-\nu)}H_{k}^{\nu-1/2}k^{-\nu}
\left[\frac{e^{-2\nu H_{k}t}}{H_{k}^{2\nu}}k^{2\nu}
\Gamma(-\nu)+4^{\nu}e^{i\nu\pi}\Gamma(\nu)\right],
\end{equation}
where we have used $a=e^{H_{k}t}$. When $\nu$ is real and positive,
this behaves as
\begin{equation}
X_{k}\approx-\frac{1}{\sqrt{\pi}}(-1)^{3/4}2^{-1-\nu}e^{i\nu\pi/2}
\left(\frac{k}{aH_{k}}\right)^{-\nu}\frac{4^{\nu}}{\sqrt{H_{k}a^{3}}}
\Gamma(\nu),
\label{eq:realnu}
\end{equation}
which can become large when $k/(aH_{k})\ll1$ and $\nu\neq0$. When
$\nu$ is imaginary and positive, the wave function behaves as
\begin{equation}
X_{k}\approx\frac{-1}{\sqrt{\pi}}(-1)^{3/4}2^{-1-|\nu|i}
e^{|\nu|\pi/2}\left(\frac{k}{aH_{k}}\right)^{i|\nu|}\frac{1}{\sqrt{H_{k}a^{3}}}
\Gamma(-i|\nu|),
\label{eq:imaginarynu}
\end{equation}
whose magnitude is essentially independent of $k$, as it can be
understood by noting that this limit corresponds to  $m\gg k$.

Let's see how the mode function $X_{k}$ scales after inflation ends.
Assuming that the scale factor scales as
\begin{equation}
a=a_{i}(t/t_{i})^{q},
\end{equation}
we find for the mode equation
\begin{equation}
\ddot{f}+\left[m^{2}+\left(\frac{3}{2q}-\frac{9}{4}\right)H^{2}(t)
+\left(\frac{k}{a(t)}\right)^{2}\right]f=0,
\end{equation}
where $X_{k}(t)=a^{-3/2}f(t)$. Neglecting the $k$ term in the
long-wavelength limit and defining $\xi_{q}\equiv3/2q-9/4$,
we find that in the limit $m^{2}\gg H^{2}$,
\begin{equation}
X_{k}\sim\frac{c}{a^{3/2}}e^{-imt},
\end{equation}
while in the limit $m^{2}\ll H^{2}$ that
\begin{equation}
X_{k}\sim\frac{c}{a^{3/2}}t^{(1+\sqrt{1-4q^{2}\xi_{q}})/2}.
\end{equation}
Hence, we see that since $H\sim t^{-1}$, eventually, $m^{2}\gg H^{2}$
and $X_{k}$ will fall like $a^{3/2}$. Hence, for any quantity we
compute involving $X_{k}$ mode function in dS space, as far as the
scaling with $a$ is concerned in the absence of further interactions,
we should freeze its value at the point when $m=H$ and then scale
it as $a^{-3/2}$.

Because of the $k$ suppression, the term
$a^{-2}(|\vec{k}_{1}|^{2}-|\vec{k}_{1}||\vec{k}|\cos\theta)
X_{k_{1}}X_{\sqrt{k^{2}+k_{1}^{2}-2k_{1}k\cos\theta}}$
in Eq.\ (\ref{eq:leadingpower}) should be negligible unless there
is a cancellation of the other terms in the limit that $|\vec{k}|\rightarrow0$.
Quantitatively, the ratio
\begin{equation}
\lim_{k,k_{1}\rightarrow0}\frac{\dot{X}_{k_{1}}
\dot{X}_{\sqrt{k^{2}+k_{1}^{2}-2k_{1}k\cos\theta}}}
{m^{2}X_{k_{1}}X_{\sqrt{k^{2}+k_{1}^{2}-2k_{1}k\cos\theta}}}
=\frac{9H^{2}}{4m^{2}}.
\end{equation}
Hence, as long as $m^2 \neq 9H^2/4$, there should be no
cancellation. Hence, the power spectrum Eq.\ (\ref{eq:leadingpower})
becomes
\begin{eqnarray}
P_{\delta_{X}} & \approx & \frac{1}{\rho_{X}^{2}}\frac{k^{3}}{2\pi}
\frac{1}{(2\pi)^{3}}\int dk_{1}k_{1}^{2}\int_{-1}^{1}d\cos\theta
\left|\dot{X}_{k_{1}}\dot{X}_{\sqrt{k^{2}+k_{1}^{2}-2k_{1}k\cos\theta}}
+m^{2}X_{k_{1}}X_{\sqrt{k^{2}+k_{1}^{2}-2k_{1}k\cos\theta}}\right|^{2}
\nonumber \\
& \approx & \frac{1}{\rho_{X}^{2}}\frac{k^{3}}{2\pi}\frac{1}{(2\pi)^{3}}
\int dk_{1}k_{1}^{2}\int_{-1}^{1}d\cos\theta\left\{\left|\dot{X}_{k_{1}}
\dot{X}_{\sqrt{k^{2}+k_{1}^{2}-2k_{1}k\cos\theta}}\right|^{2}+m^{4}\left|
X_{k_{1}}X_{\sqrt{k^{2}+k_{1}^{2}-2k_{1}k\cos\theta}}\right|^{2}\right.
\nonumber \\
 &  & \left.+ 2m^{2}\textrm{Re}\left[\dot{X}_{k_{1}}^{*}
 \dot{X}_{\sqrt{k^{2}+k_{1}^{2}
 -2k_{1}k\cos\theta}}^{*}X_{k_{1}}
 X_{\sqrt{k^{2}+k_{1}^{2}-2k_{1}k\cos\theta}}\right]\right\}.
 \label{eq:massaugedlead}
 \end{eqnarray}
In the language of \cite{Liddle:1999pr}, we have
\begin{equation}
P_{X}(k)=\frac{k^{3}}{2\pi^{2}}|X_{k}|^{2},
\end{equation}
which implies, keeping only the term proportional to $m^{4}$, that
\begin{equation}
P_{\delta_{X}}(k) = \frac{m^{4}}{\rho_{X}^{2}}\frac{k^{3}}{2\pi}
\frac{1}{(2\pi)^{3}}\int dk_{1}k_{1}^{2}\int_{-1}^{1}dxP_{X}(k_{1})
P_{X}(|\vec{k}_{1}-\vec{k}|)\frac{2\pi^{2}}{k_{1}^{3}}
\frac{2\pi^{2}}{|\vec{k}_{1}-\vec{k}|^{3}}
 = \frac{m^{4}}{\rho_{X}^{2}}k^{3}\frac{1}{8\pi}
 \int d^{3}k_{1}\frac{P_{X}(k_{1})P_{X}(|\vec{k}_{1}-\vec{k}|)}
 {k_{1}^{3}|\vec{k}_{1}-\vec{k}|^{3}}\label{Liddle}
\end{equation}
which matches the result of Ref.\ \cite{Liddle:1999pr}. However, unlike their
analysis, we do not neglect terms involving $\dot{X}_{k_{1}}
\dot{X}_{\sqrt{k^{2}+k_{1}^{2}-2k_{1}k\cos\theta}}$
which can be a priori of same order of magnitude as the
$m^{2}X_{k_{1}}X_{\sqrt{k^{2}+k_{1}^{2}-2k_{1}k\cos\theta}}$
term they accounted for. We will see the resulting correction is close
to a factor of 2.

We can express the isocurvature power spectrum Eq.\ (\ref{eq:massaugedlead})
in terms of the power spectrum of the dark matter field as follows:
\begin{eqnarray}
P_{\delta_{X}}(k) & = &
\frac{k^{3}}{\rho_{X}^{2}}\frac{1}{8\pi}\int
d^{3}k_{1}\frac{1}{k_{1}^{3}|\vec{k}-\vec{k}_{1}|^{3}} \left\{
\frac{\left(\left[\dot{P}_{X}(k_{1})\right]^{2}
+ k_{1}^{6}/a^{6}\left(2\pi^{2}\right)^{2} \right)
\left(\left[\dot{P}_{X}(|\vec{k}_{1}-\vec{k}|)\right]^{2}
+|\vec{k}-\vec{k}_{1}|^{6}/a^{6}\left(2\pi^{2}\right)^{2}\right)}
{16P_{X}(k_{1})P_{X} (|\vec{k}-\vec{k}_{1}|)}\right.\nonumber \\
&  & \left. \phantom{\frac{\left(\left[\dot{P}_{X}(k_{1})\right]^{2}\right)}
{P_{X} (|\vec{k}-\vec{k}_{1}|)}}
+m^{4}P_{X}(k_{1})P_{X}(|\vec{k}_{1}-\vec{k}|)
+\frac{m^{2}}{2}\left(\dot{P}_{X}(k_{1})\dot{P}_{X}(|\vec{k}-\vec{k}_{1}|)
-k_{1}^{3}|\vec{k}-\vec{k}_{1}|^{3}/a^{6}(2\pi^{2})^{2}\right)\right\}.
 \label{eq:dummy1}
\end{eqnarray}
Note the presence of additional terms with respect to Eq.\
(\ref{Liddle}) due to the kinetic contributions to the stress
tensor appear as derivatives of the power spectrum.

At the end of inflation, the Hubble expansion rate is much smaller
than the value at which the isocurvature perturbations are
generated. For example, in the chaotic inflationary scenario that we
consider, we have
\begin{equation} 
H_{e}\approx H_{I}/\sqrt{50} ,
\end{equation}
where $H_{I}$ is the expansion rate during the time that the
isocurvature perturbations are generated. Since gravitational
particle production is determined by
\begin{equation}
\frac{m_{X}}{H_{e}}\approx\frac{m_{X}\sqrt{50}}{H_{I}} ,
\end{equation}
and since to have enough particle production for dark matter with
$T_{RH} \lesssim 10^9$ GeV,\footnote{We will see later that the
  isocurvature bound of interest is within the regime where
  $T_{RH}<10^9$ GeV.} we must satisfy the condition
\begin{equation}
\frac{m_{X}}{H_{e}}\lesssim 7,
\end{equation} 
we must
have
\begin{equation}
\frac{m_{X}}{H_{I}}\lesssim 1.
\end{equation}
This implies that $\nu$ is real and positive, \footnote{With
nonminimal coupling, $\nu$ may easily be imaginary in which case
Eq.~(\ref{eq:imaginarynu}) will be relevant instead of
Eq.~(\ref{eq:realnu}).  In that case, the spectral index for the
isocurvature perturbations will not be as sensitive to the mass $m_X$
since $|\Gamma(-i x)| \exp(x \pi/2)$ scales as $1/\sqrt{x}$ for real $x>1$. } which
means that the wave function will behave as
\begin{equation}
X_{k}\approx -\frac{1}{\sqrt{\pi}}(-1)^{3/4}2^{-1-\nu}e^{i\nu\pi/2}
\left(\frac{k}{aH_{k}}\right)^{-\nu}
\frac{4^{\nu}}{\sqrt{H_{k}a^{3}}}\Gamma(\nu),
\label{eq:modefornureal}
\end{equation}
and the spectrum
\begin{equation}
P_{X}(k) \approx \frac{k^{3}}{2\pi^{2}}\left|\frac{-1}{\sqrt{\pi}}(-1)^{3/4}
2^{-1-\nu}e^{i\nu\pi/2} \left(\frac{k}{aH_{k}}\right)^{-\nu}\frac{4^{\nu}}
{\sqrt{H_{k}a^{3}}}\Gamma(\nu)\right|^{2}\\
= A\frac{k^{3}}{H_{k}a^{3}}\left(\frac{k}{aH_{k}}\right)^{-2\nu}
\end{equation}
where
\begin{equation}
A\equiv\frac{2^{2\nu-3}}{\pi^{3}}|\Gamma(\nu)|^{2}.
\end{equation}
When we take the time derivative of $P_{X}(k)$, we will only
account for the time dependence in $a$ since that is the main time
dependence information contained in the approximate Bessel
function equation. Hence, we find
\begin{equation}
\dot{P}_{X} \approx 
A\frac{k^{3}}{H_{k}a^{3}}\left(\frac{k}{aH_{k}}\right)^{-2\nu}(2\nu-3)H_{k}
\end{equation}
where in the second equality, we have assumed approximate dS expansion again.

Finally, using Eq.\ (\ref{eq:hkasfuncofhi}), we
find
\begin{eqnarray} 
P_{X}(k) & = &
A\frac{k^{3}}{H_{I}a^{3}}\left(\frac{k}{aH_{I}}\right)^{n}
\nonumber \\
\dot{P}_{X} & = &
A\frac{k^{3}}{a^{3}}\left(\frac{k}{aH_{I}}\right)^{n-\epsilon}(2\nu-3)
\end{eqnarray}
where
\begin{equation}
n\equiv\epsilon-2\nu(1+\epsilon).
\label{eq:powerlawindex}
\end{equation}

Hence, we can write the simplified form of the power spectrum as
\begin{eqnarray}
P_{\delta_{X}}(k) & = & \frac{k^{6}}{\rho_{X}^{2}}\frac{1}{4}\int_{0}^{\infty}
duu^{2}\left\{ \frac{k}{256a^{7}A^{2}\pi^{8}}H_{I}
\left(\frac{k}{aH_{I}}\right)^{-(1+2n)}(u)^{-(1+n)}\left(1+4A^{2}
\left(\frac{k}{aH_{I}}\right)^{2(n-\epsilon)}u^{2(n-\epsilon)}(3-2\nu)^{2}
\pi^{4}\right)\right.\nonumber \\
 &  & \times\left[|1-u|^{-n+2}\left(1+\frac{-4A^{2}(3-2\nu)^{2}\pi^{4}}
 {2(1-\epsilon)+n}\left(\frac{k}{aH_{I}}\right)^{2(n-\epsilon)}
 |1-u|^{2(n-\epsilon)}\right)\right.\nonumber \\
 &  & \left.+|1+u|^{2-n}\left(\frac{1}{2-n}+\frac{4A^{2}(3-2\nu)^{2}\pi^{4}}
 {2(1-\epsilon)+n}\left(\frac{k}{aH_{I}}\right)^{2(n-\epsilon)}
 |1+u|^{2(n-\epsilon)}\right)\right]\nonumber \\
 &  & +A^{2}\left(\frac{k}{aH_{I}}\right)^{2n-1}u^{n-1}
 \frac{m^{4}k}{a^{7}H_{I}^{3}(n+2)}\left[|1+u|^{n+2}-|1-u|^{n+2}\right]
 \nonumber \\
 &  & + \frac{m^{2}}{4a^{6}}\left[\frac{-1}{\pi^{4}}
 +\frac{-2A^{2}(3-2\nu)^{2}}{u(2-\epsilon+n)}
 \left(\frac{k}{aH_{I}}\right)^{2n-2\epsilon}|1+u|^{-\epsilon}
 |1-u|^{-\epsilon}u^{n-\epsilon}\right.
 \nonumber \\
 &  & \times\left.\left. \phantom{\left(\frac{k}{aH_{I}}\right)^{2n-2\epsilon}}
 \hspace*{-60pt}
 \left(|1-u|^{n+2}|1+u|^{\epsilon}-|1-u|^{\epsilon}
 |1+u|^{n+2}\right)\right]\right\}.
 \label{eq:powerintermed}
\end{eqnarray}
One can easily check that
the term proportional to $m^{4}$ matches the integral expression
of \cite{Liddle:1999pr}.

\begin{table}
\caption{\label{tbl:Coefficients}Coefficients to the integrals governing
the order of magnitude of the power spectrum is presented. The estimate
in the third column includes the contribution from the integral multiplying
the coefficients in the expansion of the power spectrum. As explained
in the text, the divergent integrals were cut off at $k_{\textrm{max}}=
a_{e}H_{e}$
and the $k$ value of the spectrum set at $k=a_{i}H_{I}y$ where $a_{i}$
is the scale factor at the {}``beginning'' (near 50 efolds from
the end) of the inflation and $a_{e}$ is the scale factor at the
end of inflation. The degree of divergence in the integrals is given
in the fourth column. It is clear that $a_{i}I_{i}$ contributions
to the power spectrum dominate because these terms are not diluted
by the enormous scale factor. For $c_{2}I_{2}$ (where $I_{2}$ is
the integral), the divergence is one degree smaller than the degree
of divergence in each term in the integrand because of cancellation. }
\begin{ruledtabular}
\begin{tabular}{lccr}
 &  & \textrm{order of magnitude} & \textrm{degree of } \\ 
\textrm{coeff.} & \textrm{value} & 
\textrm{contribution including integral} & \textrm{divergence}\\
\hline
$c_1$ & $(256A^{2}\pi^{8})^{-1}(k/aH_I)^{12-2n_l}$ 
  & $6\times10^{-4}(a_e/a)^{10}(a_i/a)^{2}(H_e/H_I)^{10}y^{2}$ & 10 \\
$c_2$ & $ 4(3-2\nu)^{2}\pi^{4}(n_{l}-1-\epsilon)^{-1}(256\pi^{8})^{-1}
(k/aH_I)^{6-2\epsilon}$ & $7\times10^{-5}(a_i/a)^3(a_e/a)^3
 (H_e/H_I)^3(m_X/H_I)^4y^3$  & 3 \\
$c_3$ & $c_1/(5-n_l)$ & $6\times10^{-4}(a_e/a)^{10}(a_i/a)^2(H_e/H_I)^{10}y^2$
   & 10 \\
$c_4$ & $(n_{l}-1-\epsilon)c_{2}$ &
$-7\times10^{-5}(a_i/a)^4(a_e/a)^2(H_e/H_I)^2(m_X/H_I)^4y^4$ & 2 \\
$c_5$ & $-(3-2\nu)^2(64\pi^4)^{-1}(n_l-5))^{-1}(k/aH_I)^{6-2\epsilon}$ & 
 $10^{-5}(a_i/a)^4(a_e/a)^2(H_e/H_I)^2(m_X/H_I)^4y^4 $ & 2 \\
$c_6$ & $-m_X^2(4\pi^4H_I^2)^{-1}(k/aH_I)^6$ &
 $-3\times10^{-3}(a_e/a)^3(a_i/a)^3(H_e/H_I)^3(m_X/H_I)^2y^3$ & 3 \\
$a_1$ & $A^2(3-2\nu)^416^{-1}(1+2\epsilon-n_l)^{-1}(k/aH_I)^{2n_l-4\epsilon}$ &
 $8\times10^{-6}(m_X/H_I)^8$ & \\
 $a_2$ & $A^2(m_X/H_I)^4(1-n_l)^{-1}(k/aH_I)^{2n_l}$ & 
 $6\times10^{-4}(m_X/H_I)^4$ & \\
$a_3$ & $A^2(3-2\nu)^2(m_X/H_I)^22^{-1}
   (1-n_l+\epsilon)^{-1}(k/aH_I)^{2(n_l-\epsilon)}$ & 
   $10^{-4}(m_X/H_I)^6$ & \\
\end{tabular}
\end{ruledtabular}
\end{table}

\begin{table}
\caption{\label{tbl:integrand}Integrals multiplying the coefficients in Table
\ref{tbl:Coefficients}. }
\begin{ruledtabular}
\begin{tabular}{lr}
 & integrand for the measure $\int_{0}^{k_\textrm{max}}du$ \\ \hline
$I_1$ & $u^{4-n_{l}}|1-u|^{5-n_{l}}4$ \\
$I_2$ & $u^{4-n_{l}}(|1+u|^{n_{l}-1-2\epsilon}-|1-u|^{n_{l}-1-2\epsilon})$ \\
$I_3$ & $u^{4-n_{l}}|1+u|^{5-n_{l}}$ \\
$I_4$ & $u^{2(n_{l}-\epsilon-2)}|1-u|^{5-n_{l}}4$ \\
$I_5$ & $u^{2(n_{l}-2-\epsilon)}|1+u|^{5-n_{l}}4$ \\
$I_6$ & $u^{2}$ \\
$F_1$ & $u^{n_{l}-2-2\epsilon}(|1-u|^{n_{l}-1-2\epsilon}
           -|1+u|^{n_{l}-1-2\epsilon})$ \\
$F_2$ & $u^{n_{l}-2}(|1-u|^{n_{l}-1}-|1+u|^{n_{l}-1})$\\
$F_3$ & $u^{n_{l}-2-\epsilon}|1+u|^{-\epsilon}|1-u|^{-\epsilon}
  (|1-u|^{n_{l}-1}|1+u|^{\epsilon}-|1-u|^{\epsilon}|1+u|^{n_{l}-1})$ \\
\end{tabular}
\end{ruledtabular}
\end{table}

Since we know from Eq.\ (\ref{eq:powerlawindex}) that the powerlaw
index $n$ is generically a negative number close to $-3$ (since
$m/H_{I}$ is expected to be much less than 1), we will define a
new power law index
\begin{equation} n_{l}=n+3.
\end{equation} 
The $u$ integrals can be separated as 
\begin{equation}
P_{\delta_{X}}=\frac{H_{I}^{8}}{4\rho_{X}^{2}}\left(\sum_{j=1}^{3}a_{j}F_{j}
+\sum_{i=1}^{6}c_{i}I_{i}\right), 
\end{equation}
where all the coefficients are shown in Table
\ref{tbl:Coefficients} and the integrands to the integrals are
shown in Table \ref{tbl:integrand}. The estimate in the third
column is given with any divergent integrals cut off at
$k_{\textrm{max}}=a_{e}H_{e}$ corresponding to
$u_{\textrm{max}}=a_{e}H_{e}/k$ and $k=a_{i}H_{I}y$ where
$y<10^{4}$ is a scaling parameter relevant for the CMB: \textit{e.g.,}
\begin{equation}
I_{2}=\int_{0}^{a_{e}H_{e}/k}du \ \left[u^{4-n_{l}}\left(
|1+u|^{n_{l}-1-2\epsilon} -|1-u|^{n_{l}-1-2\epsilon}\right)\right].
\end{equation}
This is a sensible cutoff since the power law form of the wave
function and the decoherence of quantum fluctuations no longer
holds when $k>aH$, and the shortest wavelength that gets stretched
beyond the horizon during inflation is $k=a_{e}H_{e}$. It is clear
that $a_{i}I_{i}$ contributions to the power spectrum dominate
because these terms are not diluted by the enormous scale factor
($a_{e}/a_{i}\gtrsim10^{22}$). This means that the prediction is
insensitive to the cutoff and the prediction is robust. Note that
in making the order of magnitude estimate in the the third column,
we have assumed that $n_{l}\sim O(m_{X}^{2}/H_{I}^{2})$.

\begin{figure}
\begin{center}
\includegraphics[scale=.75]{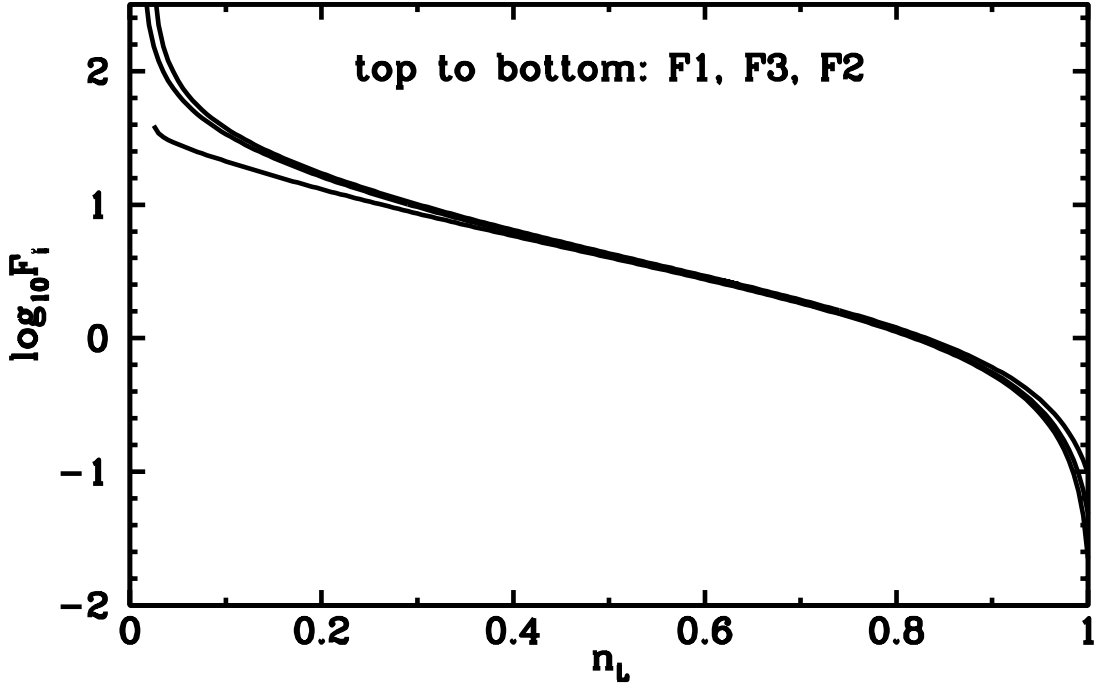}
\end{center}
\caption{\label{fig:f1f2f3plots}Plots of the functions important for the
isocurvature spectrum. Here, $\epsilon=0.01$ for $F_{1}$ and $F_{3}$.}
\end{figure}

The spectrum can thus be written simply as\begin{eqnarray}
P_{\delta_{X}}(k) & \approx &
\frac{H_{I}^{8}}{\rho_{X}^{2}}\frac{1}{4}
\left[\frac{A^{2}(3-2\nu)^{4}}{16(1+2\epsilon-n_{l})}
\left(\frac{k}{aH_{I}}\right)^{2n_{l}-4\epsilon}F_{1}(n_{l})
+\frac{A^{2}(m_{X}/H_{I})^{4}}{1-n_{l}}
\left(\frac{k}{aH_{I}}\right)^{2n_{l}}F_{2}(n_{l})
\right.\nonumber
\\ & &
\left. + \frac{A^{2}(3-2\nu)^{2}(m_{X}/H_{I})^{2}}{2(1-n_{l}+\epsilon)}
\left(\frac{k}{aH_{I}}\right)^{2(n_{l}-\epsilon)}
F_{3}(n_{l})\right] .
\label{eq:almostfinalformhurray}
\end{eqnarray}
Strictly speaking, this expression unfortunately cannot be used
beyond the end of inflation since the wave function we have used
to compute the spectrum is no longer valid even approximately.
Indeed, as we discussed earlier, when the Hubble expansion rate
falls below the mass of the dark matter, the mode function scales
like $1/a^{3/2}$. However, in the spirit of the usual scalar
density perturbations during inflation, we will simply treat the
isocurvature perturbations as a classical fluid at the end of
inflation. Owing to Eqs.\ (\ref{eq:constancyofs}) and
(\ref{eq:sdeltaandzetarelated}), this means that for long
wavelengths $P_{\delta_{X}}$ is approximately frozen at the end of
inflation assuming that the final particle production density is
simply scaled back to the end of inflation.\footnote{As discussed
before, note that we are also effectively neglecting
Bogoliubov mixing effects for the computation of the two point
function. This is also not rigorously justified in this paper but is
consistent with the usual assumptions made in computing scalar
perturbations in inflation.} Hence, our final expression for the
spectrum is
\begin{eqnarray} P_{\delta_{X}}(k) & = &
\frac{H_{I}^{8}}{\rho_{X}^{2}(t_{e})}\frac{A^{2}}{4}
\left(\frac{k}{a_{e}H_{I}}\right)^{2n_{l}}
  \left\{\frac{(3-2\nu)^{4}}{16(1+2\epsilon-n_{l})}
  \left(\frac{k}{a_{e}H_{I}}\right)^{-4\epsilon}F_{1}(n_{l})
  +\frac{(m_{X}/H_{I})^{4}}{1-n_{l}}F_{2}(n_{l})+\right.
  \nonumber  \\ & &
  \left.\frac{(3-2\nu)^{2}(m_{X}/H_{I})^{2}}{2(1-n_{l}+\epsilon)}
  \left(\frac{k}{a_{e}H_{I}}\right)^{-2\epsilon}F_{3}(n_{l})
  \right\}\label{eq:finalexpressionformally}
\end{eqnarray}
The three function $F_{i}(n_{l})$ are shown in Figure \ref{fig:f1f2f3plots}.

\section{CMB spectrum}

In this section we compute numerical values for $C_{l}$ contribution
coming from the superheavy dark matter isocurvature perturbations
within the context of a specific slow roll inflationary scenario,
namely that of $V(\phi)= \frac{1}{2} m_\phi^2 \phi^2$ scenario.  In
particular, using Eq.\ (\ref{eq:bettertotpowerspecform}), we define
the isocurvature contribution to Eq.\ (\ref{eq:finalcl}) as
\begin{equation}
C_{l}^{(X)}=\frac{16\pi}{25}\int\frac{dk}{k}P_{\delta_X}(k)[j_{l}(kL)]^{2}.
\end{equation}
For $\Omega_{\Lambda}=0$, $\Omega_{m}=1$, and $z_{dec}=\infty$,
$L$ defined by Eq.\ (\ref{eq:defofdisttols}) becomes the well known
expression $2/(a_{0}H_{0})$. Numerically, we will instead take more
realistic values of $\Omega_{\Lambda}=0.7$, $\Omega_{m}=0.24$, and
$z_{dec}=1100$, which yields
\begin{equation}
L\approx\frac{3.5}{a_{0}H_{0}}.
\end{equation}
Now, we will use the approximation
\begin{equation}
n_{l}\approx3+\epsilon-2(1+\epsilon)\nu,
\end{equation}
where $\nu$ is defined as [\textit{c.f.,} Eq.\ (\ref{eq:nufirst})]
\begin{equation}
\nu=\sqrt{\frac{9}{4}-\left(\frac{m}{H_{I}}\right)^{2}}
\end{equation}
and the slow-roll parameter is evaluated at when the longest
wavelengths of interest is leaving the horizon and $H_{I}$
corresponds to the Hubble expansion rate at that same time. Using
the well known formula
\begin{equation}
\int\frac{dk}{k}\left(\frac{k}{a_{e}H_{I}}\right)^{q}[j_{l}(kL)]^{2} = 
\frac{2^{q-3}\pi}{(a_{e}H_{I}L)^{q}}
\frac{\Gamma(l+q/2)\Gamma(2-q)}{\Gamma(l+2-q/2)\Gamma^{2}(3/2-q/2)},
\label{eq:gammaidentities}
\end{equation}
we find
\begin{eqnarray}
C_{l}^{(X)} & = & \frac{16}{25}\frac{H_{I}^{8}}{\rho_{X}^{2}(t_{e})}
\frac{1}{4}\frac{2^{4\nu-6}}{\pi^{5}}|\Gamma(\nu)|^{4}
\left[\frac{(3-2\nu)^{4}}{16(1+2\epsilon-n_{l})}F_{1}(n_{l})
\frac{2^{2n_{l}-4\epsilon-3}\pi}{(a_{e}H_{I}L)^{2n_{l}-4\epsilon}}
\frac{\Gamma(l+n_{l}-2\epsilon)\Gamma(2-2n_{l}+4\epsilon)}
{\Gamma(l+2-n_{l}+2\epsilon)\Gamma^{2}(3/2-n_{l}+2\epsilon)}\right.
\nonumber \\
 &  & + \frac{(m_{X}/H_{I})^{4}}{1-n_{l}}F_{2}(n_{l})
 \frac{2^{2n_{l}-3}\pi}{(a_{e}H_{I}L)^{2n_{l}}}
 \frac{\Gamma(l+n_{l})\Gamma(2-2n_{l})}
 {\Gamma(l+2-n_{l})\Gamma^{2}(3/2-n_{l})}\nonumber \\
 &  & + \left. \frac{(3-2\nu)^{2}(m_{X}/H_{I})^{2}}{2(1-n_{l}+\epsilon)}
 F_{3}(n_{l})\frac{2^{2n_{l}-2\epsilon-3}\pi}{(a_{e}H_{I}L)^{2n_{l}-2\epsilon}}
 \frac{\Gamma(l+n_{l}-\epsilon)\Gamma(2-2n_{l}+2\epsilon)}
 {\Gamma(l+2-n_{l}+\epsilon)\Gamma^{2}(3/2-n_{l}+\epsilon)}\right]
 \label{eq:finalform}
\end{eqnarray}
Note that Eq.\ (\ref{eq:gammaidentities}) says that the effect of the
Bessel function integral is to merely set $k$ to $1/L$ and multiply
the $(k/(a_e H_I))^q|_{k=1/L}$ by a dimensionless number of order $0.1$.

Finally, for the numerical analysis, we need an expression for
$\rho_{e}$. As we explained in the appendix, the infrared
contribution to this density is ambiguous. Although an accurate
computation of the non-infrared contribution can only be computed
on a model by model basis, the mass bound coming from the
isocurvature contribution is not sensitive to the uncertainty
because of the large number in the exponential. On the other hand,
the  uncertainty on the reheating temperature is the same as the
one in the density of particles produced. Hence, the reheating
temperature bound is uncertain by a factor of about 10.

\begin{figure}
\begin{center}
\includegraphics[scale=.75]{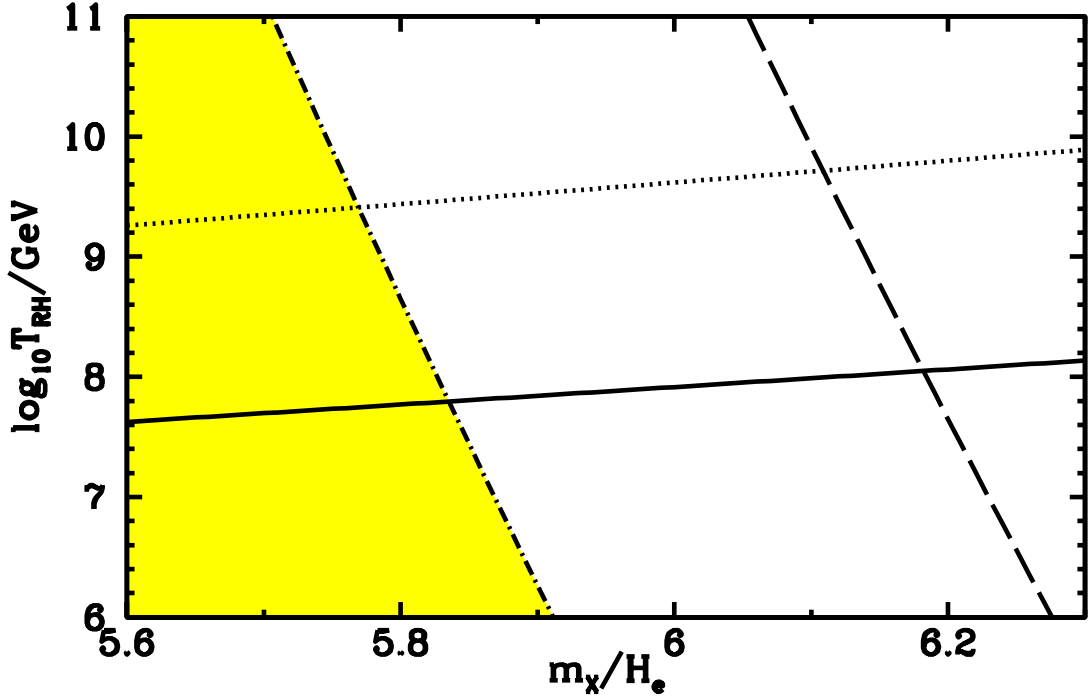}
\end{center}
\caption{\label{fig:constraint}The region of parameter space excluded
by the isocurvature constraint. The shaded area bounded by dot-dashed
curve is ruled out due to overproduction of isocurvature
perturbations.  The bound on $C_{l}^{(X)}$ we took was
$5\times10^{-10}$ corresponding to
$C_{l}l(l+1)/(2\pi)|_{l=2}=3532\,(\mu K)^{2}$.  The dashed curve
indicates the uncertainty in the isocurvature computation due to the
uncertainty in the particle production computation. Below the solid
curve, there is not enough gravitational production of particles to
have sufficient dark matter ($\Omega_{X}h^{2}<0.1$). Above the dotted
curve, there is overproduction ($\Omega_{X}h^{2}>0.2$) of dark
matter. Although this specific example is for
$V=\frac{1}{2}m_{\phi}^{2}\phi^{2}$ inflationary potential, as we
argued in Section 2, the constraints are general in that superheavy
dark matter scenario requires a reheating temperature of above
$10^{8}$ GeV.}
\end{figure}

For the slow-roll inflationary scenario of
$V_{\phi}=\frac{1}{2}m_{\phi}^{2}\phi^{2}$, one can compute the
non-infrared contributions to the $\rho_{X}(t_{e})$ to be bounded
as shown in Eq.\ (\ref{eq:quadraticinflationbound}). For a more
general class of slow roll inflationary models, the particle
energy density can be estimated by Eq.\ (\ref{eq:generalized}). To
obtain the numerical value of $a_{e}H_{I}L$ appearing in Eq.\
(\ref{eq:finalform}), we use the standard reheating
relationships
\begin{eqnarray}
\frac{a_{e}}{a_{RH}} & = & \left(\frac{\rho_{RH}}{\rho_{e}}\right)^{1/3} =
(2.2\times10^{-62})^{1/3}\left(\frac{T_{RH}}{\mathrm{GeV}}\right)^{4/3} 
\nonumber\\  
\frac{a_{RH}}{a_{0}} & = & 
\left(\frac{g_{*S}(t_{0})}{g_{*S}(t_{RH})}\right)^{1/3}\frac{T_{0}}{T_{RH}}=
8.1\times10^{-14}\frac{\mathrm{GeV}}{T_{RH}}
\end{eqnarray}
In Fig.\ \ref{fig:constraint} we plot the parameter space allowed
by the combined constraints of isocurvature fluctuations and
$\Omega_{CDM}$ in the $V=\frac{1}{2}m_{\phi}^{2}\phi^{2}$ slow
roll inflationary model.  The somewhat arbitrary bound on
$C_{l}^{(X)}$ we took was $5\times10^{-10}$ corresponding to
$C_{l}l(l+1)/(2\pi)|_{l=2}=3532\,(\mu K)^{2}$ in units of
temperature. This rather large upper bound was taken to obtain a
conservative bound on the mass of the superheavy dark matter.
Because of the largeness of $a_e H_I L$ in Eq.\
(\ref{eq:finalform}), the mass bound is not sensitive to the exact
bound of $C_l$ we use.  As one can see, since the dark matter must
compose the CDM of the universe, we have a robust bound on the
reheating temperature of
\begin{equation} T_{RH}>10^{8}\textrm{GeV}.
\end{equation} and on the mass of the superheavy dark matter
of about
\begin{equation}
\frac{m_{X}}{H_{e}}\gtrsim 6.
\end{equation} 
As argued in Sec.\ 2,
this bound on the mass and the reheating temperature should be
robust for most slow roll inflationary models. The rheating
temperature bound has an uncertainty of about a factor of $10$ and
the $m_{X}/H_{e}$ bound has an uncertainty of less than unity.
Note that one cannot naively extrapolate using our results to
situations to large $m_{X}/H_{e}$ (corresponding to small lower
bound on reheating temperature) because there, the gravitational
production is insufficient to compose the dark matter and the
computations have assumed that the gravitationally produced
superheavy dark matter dominates the matter density during matter
domination. When the gravitationally produced contribution becomes
negligible fraction of CDM, we expect the isocurvature bounds to
disappear. \footnote{It is important to remember that this work
has been done assuming that the superheavy dark matter makes up
all of the dark matter.}  Unlike in section 2, the tensor to
scalar amplitude limit does not play much role here because the
tensor to scalar ratio here is predicted in this specific
inflationary scenario to be much less than unity (of order $10
\epsilon \approx 0.1$) on long wavelengths, consistent with
$r_S=0.1$. It is important to note that all of the results in this
section are in agreement with the more general discussion of
Sec.\ 2.

One might worry that since $m_{X}/H_{e}$ considered in Fig.\
\ref{fig:constraint}
corresponds to a number larger than 1, we may have violated our assumption
that $\nu$ is real.
In most inflationary scenarios, this constraint is not violated since
$H_{I}\gtrsim3H_{e}$ even when slowly rolling. (For example, $H_{I}>7H_{e}$
for $V=\frac{1}{2}m_{\phi}^{2}\phi^{2}$ type potential.)

Note also that numerically, the relative contributions to $C_{l}^{(X)}$
of the $\{ F_{1},F_{2},F_{3}\}$ terms are
\begin{equation}
\textrm{(term }F_{1}):(\textrm{term }F_{2}):(\textrm{term }F_{3})
=1:11:7
\end{equation}
at $m_{X}/H_{e}=6$ and $T_{RH}=10^{9}$. These ratios are typical
in the region of interest. Hence, the kinetic contribution gives almost
a factor of 2 correction to the isocurvature perturbation computation.

\section{Conclusion}

Isocurvature perturbations are generic prediction of inflationary
cosmology. We have explored the isocurvature perturbation constraint
on models of superheavy dark matter minimally coupled to
gravity. Surprisingly, there is a robust lower bound on the reheating
temperature for these models of about $10^{7} (0.2/r_S)$ GeV where
$r_S$ is the bound on tensor to scalar. This means that, if the
superheavy dark matter scenario is embedded in supergravity models
with gravity mediated SUSY breaking, the gravitino bound, which is
typically an upper bound on the reheating temperature of about
$10^{8}$ GeV, strongly squeezes these models.  If, for example, LHC
data favors a gravity mediated SUSY breaking scenario in which the
cascade of gravitino decay reactions favors an upper bound on the
reheating temperature of $10^{7}$ GeV, then superheavy dark matter
scenario can be ruled out.  Note also that because of the presence of
tensor to scalar power limit $r_S$, improved experimental bounds on
this quantity can also squeeze this dark matter scenario.

There is also a corresponding $m_{X}/H_{e}>\mathcal{O}(5)$ bound
on the mass $m_{X}$ of the superheavy dark matter. This implies
that the gravitational particle production is in the exponentially
suppressed regime since the exponent of the suppression is of
order $-2m_{X}/H_{e}$: i.e. noninfrared modes make up most of the
dark matter. Moreover, as can be seen in Eq.\
(\ref{eq:helowerbound}), one cannot lower $H_{e}$ to less than
$3\times 10^{12}$GeV while keeping $m_{X}/H_{e}$ fixed and
$T_{RH}<10^9$GeV to obtain sufficient abundance of dark matter
because the overall dark matter density scales as $m_{X}^{5/2}$.\footnote{Note 
that for $m_{X}<H_{e}$, this scaling law is
expected to change to something closer to a power $2$, but the
present argument still applies. } Hence, we see that for the
superheavy CDM scenario of gravitational particle production, CMB
data forcing the isocurvature contribution to be small favors a
relatively large gravitational wave amplitude.

\section*{Acknowledgments}
E.W.K.\ is supported in part by NASA grant NAG5-10842 and by the Department of
Energy.  D.J.H.C.~acknowledges the hospitality of Aspen Center for Physics
and CERN where parts of this work were carried out. L.S. is supported in 
part by the Department of Energy under cooperative 
research agreement DF-FC02-94ER40818, and he would like to thank R. Barbieri 
and the Scuola Normale Superiore of Pisa, and Fermilab, 
where parts of this work were carried out. 

\begin{appendix}
\section{Meaning of Isocurvature Perturbations}

Given $N$ particle species with their energy densities different
from the average energy density, the total energy densities contribute
to the curvature perturbations (because local energy density determines
the gravitational potential). The $N-1$ remaining energy density
field degrees of freedom can contribute to what is often called isocurvature
perturbations since $N-1$ degrees of freedom can be assigned any
value while keeping the the total energy density and hence the curvature
constant.

To write down explicitly the usual condition for adiabatic perturbations,
denote the energy density deviation from the average density of particle
species $i$ as
\begin{equation}
\delta_{i}=\frac{\delta\rho_{i}}{\rho_{i}}.
\end{equation}
For perturbations to be adiabatic, we have 
 \begin{equation}
\delta(s_{i}a^{3})=0
\label{eq:entropyconserve}
\end{equation}
where $s_{i}$ is the entropy density and $a$ is the scale factor.
Since the photon entropy is conserved assuming adiabatic evolution
of the photons, Eq.\ (\ref{eq:entropyconserve}) implies
\begin{equation}
\delta(s_{i}/s_{\gamma})=0.
\end{equation}
Since according to first law of thermodynamics, we have 
$s_{\gamma}=(\rho_{\gamma}+P_{\gamma})/T=\frac{4}{3}\rho_{\gamma}/T$,
we find
\begin{equation}
\frac{\delta(s_{i}/s_{\gamma})}{(s_{i}/s_{\gamma})} = 
\frac{\delta s_{i}}{s_{i}}-\frac{\delta s_{\gamma}}{s_{\gamma}}
=  \frac{\delta s_{i}}{s_{i}}-3\frac{\delta T}{T}
= \frac{\delta s_{i}}{s_{i}}-\frac{3}{4}\delta_{\gamma}=0
\label{eq:adiabaticcondition}
\end{equation}
 as the condition for adiabatic perturbations.

Now, we can derive a useful relationship in terms of equation of state.
First, the first law of thermodynamics gives
\begin{equation}
d(\rho_{i}a^{3})=T_{i}d(s_{i}a^{3})-P_{i}da^{3}+\mu_{i}d(n_{i}a^{3}).
\end{equation}
Setting the entropy to zero gives
\begin{equation}
d\rho_{i}a^{3}=-d(a^{3})[\rho_{i}+P_{i}-\mu_{i}n_{i}]+\mu_{i}dn_{i}a^{3}
\end{equation}
or equivalently
\begin{equation}
\delta_{i}=3\frac{\delta a}{a}\left[\frac{\mu_{i}n_{i}}
{\rho_{i}}-\left(1+\frac{P_{i}}{\rho_{i}}\right)\right]
+\mu_{i}\frac{\delta n_{i}}{\rho_{i}}.
\label{eq:energydensitypert}
\end{equation}
The adiabatic condition Eq.\ (\ref{eq:entropyconserve}) gives 
\begin{equation}
\frac{\delta s_{i}}{s_{i}}  =  -3\frac{\delta a}{a} = 
 \frac{\delta_{i}-\mu_{i}\delta n_{i}/\rho_{i}}
 {1+P_{i}/\rho_{i}-\mu_{i}n_{i}/\rho_{i}}
 \label{eq:entropyperturb}
\end{equation}
where we have used Eq.\ (\ref{eq:energydensitypert}). Hence, using
Eqs.\ (\ref{eq:adiabaticcondition}) and (\ref{eq:entropyperturb}),
the adiabatic condition can be written as
\begin{equation}
\frac{\delta_{i}-\mu_{i}\delta n_{i}/\rho_{i}}
{1+w_{i}-\mu_{i}n_{i}/\rho_{i}}-\frac{3}{4}\delta_{\gamma}=0.
\label{eq:consteqofstate}
\end{equation}
Note that here all evolution was adiabatic. For the nonrelativistic
particles, this expression simplifies further since
\begin{equation}
\frac{\delta_{i}-\mu_{i}\delta n_{i}/\rho_{i}}
{1+w_{i}-\mu_{i}n_{i}/\rho_{i}}-\frac{3}{4}\delta_{\gamma}=
\frac{\delta_{i}(1-\mu_{i}/m_{i})}{1-\mu_{i}/m_{i}}-\frac{3}{4}\delta_{\gamma}
=\delta_{i}-\frac{3}{4}\delta_{\gamma}=0.
\end{equation}
which is the usual condition that we see in the literature.

It is interesting to rederive Eq.\ (\ref{eq:consteqofstate}) from
a kinetic theory point of view. From a microcanonical ensemble construction,
we can write the entropy density as
\begin{equation}
s_{i}=\int\frac{d^{3}p}{(2\pi)^{3}}a^{-3}f_{i}(1-\ln f_{i}),
\label{eq:entropy}
\end{equation}
where $p$ is the comoving momentum and $f$ is the phase space distribution
function. This directly gives
\begin{equation}
\delta s_{i}=-\int\frac{d^{3}p}{(2\pi)^{3}}a^{-3}\delta f_{i}\ln f_{i}-3
\frac{\delta a}{a}s_{i}.
\label{eq:entropydev}
\end{equation}
On the other hand, the adiabaticity condition Eq.\ (\ref{eq:entropyconserve})
implies
\begin{equation}
\delta s_{i}=-3\frac{\delta a}{a}s_{i}.
\end{equation}
Hence, we are lead to conclude
\begin{equation}
\int\frac{d^{3}p}{(2\pi)^{3}}a^{-3}\delta f_{i}\ln f_{i}=0.
\label{eq:distributionconstraint}
\end{equation}
In the near equilibrium case without degeneracy, $\ln f_{i}<0$. If
in addition, if $\delta f_{i}=0$, we can conclude 
$\delta n_{i}=-3a^{-1}\delta a\int d^{3}p(2\pi)^{-3}a^{-3}f_{i}
=-3a^{-1}n_{i}\delta a .$
However, generically, the constratin is only given by 
Eq.\ (\ref{eq:distributionconstraint}).

If we assume equilibrium situation without degeneracy, we have
\begin{equation}
s_{i} =  \int\frac{d^{3}p}{(2\pi)^{3}}a^{-3}f_{i}\left(1+\frac{E_{i}-\mu_{i}}
{T_{i}}\right) = n_{i}+\frac{\rho_{i}-\mu_{i}n_{i}}{T_{i}} .
\end{equation}
The constraint Eq.\ (\ref{eq:distributionconstraint}) gives
\begin{equation}
\int\frac{d^{3}p}{(2\pi)^{3}}a^{-3}\delta f_{i}\left(\frac{E_{i}-\mu_{i}}
{T_{i}}\right)=0.
\end{equation}
Since
\begin{equation}
\rho_{i}=\int\frac{d^{3}p}{(2\pi)^{3}}a^{-3}E_{i}f_{i},
\end{equation}
we have
\begin{eqnarray}
\delta\rho_{i} & = & -3\frac{\delta a}{a}\rho_{i}
+\int\frac{d^{3}p}{(2\pi)^{3}}a^{-3}E_{i}\delta f_{i}
+\int\frac{d^{3}p}{(2\pi)^{3}}a^{-3}\delta E_{i}f_{i} \nonumber \\
 & = & -3\frac{\delta a}{a}\rho_{i}+\mu_{i}\int\frac{d^{3}p}{(2\pi)^{3}}a^{-3}
 \delta f_{i}-\frac{\delta a}{a}\int\frac{d^{3}p}{(2\pi)^{3}}a^{-3}
 \frac{|\vec{p}/a|^{2}}{E_{i}}f_{i} \nonumber \\
 & = & -3\frac{\delta a}{a}(\rho_{i}+P_{i})+\mu_{i}\int
 \frac{d^{3}p}{(2\pi)^{3}a}a^{-3}\delta f_{i}
\end{eqnarray}
where we have used the adiabaticity constraint. Also, we have
\begin{equation}
\delta n_{i}=\int\frac{d^{3}p}{(2\pi)^{3}}a^{-3}\delta f_{i}
-3\frac{\delta a}{a}n_{i}.
\end{equation}
Hence, we write
\begin{equation}
\delta\rho_{i} = -3\frac{\delta a}{a}(\rho_{i}+P_{i})
+\mu_{i}\left(\delta n_{i}+3\frac{\delta a}{a}n_{i}\right)
 =  -3\frac{\delta a}{a}(\rho_{i}+P_{i}-\mu_{i}n_{i})+\mu_{i}\delta n_{i} ,
\end{equation}
arriving at
\begin{equation}
\frac{\delta s_{i}}{s_{i}} = -3\frac{\delta a}{a}
=\frac{\delta\rho_{i}-\mu_{i}\delta n_{i}}{\rho_{i}+P_{i}-\mu_{i}n_{i}}
= \frac{\delta_{i}-\mu_{i}\delta n_{i}/\rho_{i}}
{1+w_{i}-\mu_{i}n_{i}/\rho_{i}}.
\end{equation}
This is precisely the formula that we derived with thermodynamics.

Now, we can easily show that adiabatic perturbations usually imply
nonvanishing curvature perturbations. According to 
Eq.\ (\ref{eq:energydensitypert}),
we have
\begin{equation}
\frac{\delta\rho}{\rho} = \frac{1}{\rho}\sum_{i}\rho_{i}\delta_{i}
= -3\frac{\delta a}{a}\frac{1}{\rho}\sum_{i}\rho_{i}
\left[(1+w_{i})-\frac{\mu_{i}n_{i}}{\rho_{i}}\right]
+\frac{1}{\rho}\sum_{i}\mu_{i}\delta n_{i} =  
-3\frac{\delta a}{a}\frac{1}{\rho}\sum_{i}T_{i}s_{i}
+\frac{1}{\rho}\sum_{i}\mu_{i}\delta n_{i},
\end{equation}
which is nonvanishing as long as $\mu_{i}\delta n_{i}$ term does
not cancel the entropy term. If the adiabatic perturbations conserve
particle number (which would be true for adiabatic perturbations of
nonrelativistic particles such as baryons which carry nonnegligible
chemical potential), we can write 
\begin{equation}
\delta n_{i}=-3\frac{\delta a}{a}n_{i} ,
\end{equation}
which would imply there being no cancellation and $\delta\rho/\rho\neq0$.
On the other hand, it is clear that one can have energy density perturbed
without conserving entropy locally. Hence, curvature perturbations
do not imply adiabatic perturbations.

\section{Particle production in the large mass limit}

\subsection{General consideration of the exponential suppression}

In this appendix, using \cite{Chung:1998bt}, we give an approximation
of the exponential damping of particle production in the
regime $m_{X}/H\rightarrow\infty.$

Consider the approximate $|\beta|^{2}$ formula in \cite{Chung:1998bt}
\begin{eqnarray}
|\beta_{k}|^{2} & \approx & \exp\left(-{\frac{4(k/\sqrt{C(r)})^{2}}
{m_{X}\sqrt{C''(r)/(2C^{2}(r))}}
-\frac{4m_{X}}{\sqrt{C''(r)/(2C^{2}(r))}}}\right)
\label{eq:approxbogo} \\
C(\eta) & = & a^{2}(\eta)\left[1+\left(\xi-\frac{1}{6}\right)
\frac{6a''/a^{3}}{m_{X}^{2}}\right],
\end{eqnarray}
where $r$ is the real-value solution to the equations
\begin{eqnarray}
\frac{\mu^{2}}{6}C'''(r)+C'(r) & = & 0
\label{eq:const1} \\
\frac{\mu^{2}}{2}C''(r) + \frac{k^{2}+m_{X}^{2}C(r)}{m_{X}^{2}} & = & 0 
\label{eq:const2}
\end{eqnarray}
with $\mu$ a pure imaginary number (that also needs to be solved along
with $r$). The primes are conformal time derivatives as usual. Note
that this approximation is valid only if the conditions 
\begin{eqnarray}
\left|\frac{C^{'''''}(r)}{C'''(r)}\right|
\frac{(k^{2}+m_{X}^{2}C(r))}{10m_{X}^{2}C''(r)} & \ll & 1
\label{eq:taylorvalid1} \\
\frac{|C^{''''}(r)|}{[C''(r)]^{2}}\frac{(k^{2}+m_{X}^{2}C(r))}{6m_{X}^{2}}
& \ll & 1
\label{eq:taylorvalid2} \\
\mu^{2} & < & 0
\end{eqnarray}
are valid.

To use this formula, we need to solve for $r$ given an inflationary
 model with potential $V(\phi)$ which determines $C$.  Solving
 Eqs.\ (\ref{eq:const1}) and (\ref{eq:const2}) for $r$ is equivalent to
 solving
\begin{equation}
\frac{C'(r)C''(r)}{C(r)C'''(r)}=1+\frac{k^{2}}{C(r)m_{X}^{2}} 
\label{eq:numreq}
\end{equation}
from which we see that $r$ is a $k$ dependent quantity in general.
Now, consider the large mass expansion:
\begin{equation}
 C(\eta)\approx
 a^{2}(\eta)\left[1+O(H^{2}/m_{X}^{2})\right].
\label{eq:largemass}
\end{equation}
We then write Eq.\ (\ref{eq:numreq}) as
\begin{equation}
 q\equiv\frac{2a'(a'^{2}+aa'')}{a(3a'a''+aa''')}
 =1+\frac{k^{2}}{a^{2}(r)m_{X}^{2}}
\label{eq:polelocation}
\end{equation}
where the primes are with respect to conformal time. Note that this
 equation gives $r$ as a function of $k$. Hence we will implicitly
 mean $r=r_{k}$.

Using standard equation of motion relationships one can easily 
derive\footnote{From now on, we will be using the convention
$M_{pl}^{2}/8\pi=1$.}
\begin{equation}
q = \frac{2H[-(\dot{\phi}^{2}-V)/3+2H^{2}]}
{H(-2\dot{\phi}^{2}/3+8V/3+4H^{2})+V_{,\phi}\dot{\phi}}.
\end{equation}
Now, substituting $\dot{\phi}^{2}=2[3H^{2}-V]$, we find
\begin{eqnarray}
q & = & \left[2+\frac{1}{2VH}\frac{d}{dt}V(\phi(t))\right]^{-1} ,
\end{eqnarray}
which is a remarkably simple formula.

Substituting for $\dot{\phi}$ in terms of $H$ and $V$, we find
\begin{equation}
q = \frac{2HV}{4HV+\textrm{sign}(\dot{\phi})V_{,\phi}\sqrt{2[3H^{2}-V]}} .
\end{equation}
Except for the Taylor expansion approximation and the saddle point
approximation that led to Eq.\ (\ref{eq:approxbogo}), we have only
used one additional assumption of Eq.\ (\ref{eq:largemass}) thus far.
Now, we reparameterize $r=r_{k}$ by writing 
\begin{equation}
\dot{\phi}^{2}(r)\approx\lambda V(\phi(r)) ,
\label{eq:approxguessforr}
\end{equation}
where $\lambda$ is a numerical coefficient of $O(1)$ {[}anticipating
the fact that $r$ will be near the end of inflation{]}. Again, since
$r=r_{k}$, we have $\lambda=\lambda_{k}$. The utility of this parameterization
is to bound $\lambda$ thereby bounding the final Bogoliubov coefficient.

We can then write
\begin{equation}
H^{2}\approx\frac{1}{3}\left(\frac{\lambda}{2}+1\right)V.
\end{equation}
Using $\textrm{sign}(\dot{\phi})=-\textrm{sign}(V_{,\phi})$
with $H>0$, we find
\begin{equation}
q = \left(2-\frac{1}{2}\sqrt{\frac{3\lambda}{1+\lambda/2}
\frac{|V_{,\phi}|}{V}}\right)^{-1}.
\end{equation}
Hence, we arrive at the condition determining $r$ 
[Eq.\ (\ref{eq:polelocation})]as
\begin{equation}
\left(2-\frac{1}{2}\sqrt{\frac{3\lambda}{1+\lambda/2}}
\frac{|V_{,\phi}|}{V}\right)^{-1}\approx1+\frac{k^{2}}{a^{2}(r)m_{X}^{2}} ,
\label{eq:rsimpconst}
\end{equation}
where the $k$ dependence cannot be trusted for $k/a\lesssim H$.
However, this does not matter here since we are concerned with $H\ll m_{X}$.

Now, let us evaluate
\begin{equation}
W \equiv \frac{C''}{2C^2} =  2H^{2}+\frac{\ddot{a}}{a}, 
\end{equation}
which appears in the denominator of the exponential in
Eq.\ (\ref{eq:approxbogo}).
Since Einstein's equation
with inflaton domination gives
\begin{equation}
\ddot{a}=-\frac{a}{3}(\dot{\phi}^{2}-V),
\end{equation}
we find $W  =  V$. Hence, we find the approximate Bogoliubov coefficient 
amplitude to
be given by Eq.\ (\ref{eq:approxbogo}) with $C''/2C^{2}=V$
and $r$ given by the solution to Eq.\ (\ref{eq:rsimpconst}). Explicitly,
the Bogoliubov coefficient is approximated by
\begin{equation}
|\beta_{k}|^{2}\approx 
\exp\left(-\frac{4(k/\sqrt{C(r)})^{2}}{m_{X}\sqrt{V(\phi(r))}}
-\frac{4m_{X}}{\sqrt{V(\phi(r))}}\right).
\end{equation}

\subsection{Application}

Consider the quadratic monomial inflaton potential:
\begin{equation}
V=\frac{1}{2}m_{\phi}^{2}\phi^{2}.
\end{equation}
We find
\begin{equation}
|\beta_{k}|^{2}\approx\exp\left\{-\frac{4\sqrt{2}}{\phi/\bar{M}_P}\left[
\frac{(k/a(r))^{2}}{m_{X}m_{\phi}}+\frac{m_{X}}{m_{\phi}}\right]\right\},
\end{equation}
where we have restored the reduced Planck mass. We use Eq.\
(\ref{eq:rsimpconst}) and find
\begin{equation}
\frac{\phi(r)}{\bar{M}_P}\approx \sqrt{\frac{3\lambda}{1+\lambda/2}}
\frac{k^{2}/a^{2}(r)+m_{X}^{2}}
{2k^{2}/a^{2}(r)+m_{X}^{2}}.
\label{eq:rrootcond}
\end{equation}
Hence, we arrive at the Bogoliubov coefficient
\begin{equation}
|\beta_{k}|^{2}\approx
\exp\left[-4\sqrt{\frac{2+\lambda_{k}}{3\lambda_{k}}}
\left(\frac{2k^{2}/a^{2}(r_{k})+m_{X}^{2}}{m_{\phi}m_{X}}\right)\right].
\label{eq:bogocoefficientbestest}
\end{equation}
Note the nontrivial factor of $2$ in front of the $k^{2}/a^{2}$.

Now, we have to solve for $\lambda_{k}$ in Eq.\ (\ref{eq:approxguessforr}).
Let us look for the field value at which 
\begin{equation}
\dot{\phi}^{2}(r_{k})\approx\lambda_{k}V(\phi)
=\frac{\lambda}{2}m_{\phi}^{2}\phi^{2}(r_{k}).
\end{equation}
There is no closed form solution to this. Suppose we take the standard
slow-roll-like approximation except with $H^{2}\approx2V/3$ instead
of $H^{2}\approx V/3$. We find
\begin{equation}
\phi=\sqrt{\frac{2}{3}}\frac{\bar{M}_P}{\sqrt{\lambda_{k}}}.
\label{eq:simpend}
\end{equation}
We would then write Eq.\ (\ref{eq:rrootcond}) as
\begin{equation}
\sqrt{\frac{2}{3}}\frac{1}{\sqrt{\lambda_{k}}}\approx
\sqrt{\frac{3\lambda_{k}}
{1+\lambda_{k}/2}}\frac{k^{2}/a^{2}(r_{k})+m_{X}^{2}}
{2k^{2}/a^{2}(r_{k})+m_{X}^{2}}.
\end{equation}
This can then solved for $\lambda_{k}$ as
\begin{equation}
\lambda_{k}=\frac{1+\sqrt{1+72f^{2}}}{18f^{2}}
\end{equation}
where
\begin{equation}
f\equiv\frac{k^{2}/a^{2}(r_{k})+m_{X}^{2}}{2k^{2}/a^{2}(r_{k})+m_{X}^{2}}.
\end{equation}
Now $a(r_{k})$ still needs to be determined. Since $f$ is a function
bounded by $1/2$ and $1$ (with the long-wavelength being $1$),
we have
\begin{equation}
0.53 \lesssim\lambda_{k}\lesssim 1.2
\end{equation}
where the long-wavelength limit corresponding to $\lambda_{k}\approx0.53.$
For $m_{X}\gg m_{\phi}\sim H$, we will have mostly nonrelativistic
contributing to the particle production. For $[k/a(r_{k})]^{2}=m_{X}^{2}$,
we have $\lambda_{k}\approx0.84$. Hence, a more realistic range for
$\lambda_{k}$ is
\begin{equation}
0.53\lesssim\lambda_{k} \lesssim 0.84.
\end{equation}
From now on, we will take this to be the uncertainty range of this
parameter.

This means that $a(r_{k})$ is bounded as well. Since by Eq.\ (\ref{eq:simpend}),
we have
\begin{equation}
0.89\bar{M}_P\leq\phi\leq1.1\bar{M}_P
\end{equation}
where the long-wavelength limit corresponds to $1.1\bar{M}_P$.
Since the Friedmann equations give 
\begin{equation}
\int\frac{da}{a}=\int\frac{d\phi}
{\bar{M}_P}\sqrt{\frac{1}{6}[\dot{\phi}^{2}+m_{\phi}^{2}\phi^{2}]},
\end{equation}
we can write
\begin{equation}
\frac{\Delta a}{a(r_{k_{1}})} \approx -\frac{\Delta\phi}{\bar{M}_P}
\sqrt{\frac{1}{6}}\sqrt{1+\frac{2}{\lambda_{k_{1}}}}
\end{equation}
where $\Delta a/a\equiv[a(r_{k_{2}})-a(r_{k_{1}})]/a$ and 
$\Delta\phi/\bar{M}_P\equiv[\phi(r_{k_{2}})-\phi(r_{k_{1}})]/\bar{M}_P$
are both assumed to be small quantities and $\lambda_{k_{1}}\approx0.53$
while $\lambda_{k_{2}}\approx0.84$. This results in 
\begin{equation}
\frac{\Delta a}{a(r_{k_{1}})}\approx0.21.
\end{equation}

To find density of particles produced along with the uncertainty,
we can integrate Eq.\ (\ref{eq:bogocoefficientbestest}) assuming both
$\lambda=\lambda_{k_{i}}$ and $a(r)=a(r_{k_{i}})$ are constants
independent of $k$ and then evaluating the integral for two extreme
values of $\lambda_{k_{i}}$. The integral is
\begin{equation}
n = \int\frac{dkk^{2}}{a^{3}(2\pi^{2})}|\beta_{k}|^{2} =
\left(\frac{a(r_{k_{i}})}{a(t)}\right)^{3}
\frac{(m_{\phi}m_{X})^{3/2}}
     {128\pi^2[(2+\lambda_{k_{i}})/3\lambda_{k_i}]^{3/4}}
\sqrt{\frac{\pi}{2}}
\exp \left(   -\frac{4m_X}{m_\phi}
                \sqrt{\frac{2+\lambda_{k_i}}{3\lambda_{k_i}}}    
 \right)   .
\end{equation}
Hence, we can bound the number density to be
\begin{equation}
0.7\times10^{-3}\left(\frac{a(r_{k_{1}})}{a(t)}\right)^{3}
(m_{\phi}m_{X})^{3/2}
e^{-5.0m_X/m_\phi}\lesssim n\lesssim1.6\times10^{-3}
\left(\frac{a(r_{k_{1}})}{a(t)}\right)^{3}(m_{\phi}m_{X})^{3/2}
e^{-4.2m_X/m_\phi}.
\label{eq:appendixderivationfin}
\end{equation}

Expectedly, note that the $m_{X}$ dependence for $m_{X}\rightarrow0$
of the density here differs from that of \cite{Chung:2001cb}.
The reason why this is expected is that Eq.\ (\ref{eq:largemass})
no longer applies in that limit. We use Eq.\ (\ref{eq:appendixderivationfin})
in the main body of the paper (\textit{e.g.,} 
in Eq.\ (\ref{eq:quadraticinflationbound})).

\end{appendix}




\begin{thebibliography}{999}
\frenchspacing

\bibitem{lr} D. H. Lyth and A. Riotto,
Phys.\ Rept.\  {\bf 314}, 1 (1999).
\bibitem{Peiris:2003ff}
H. V. Peiris {\it et al.},
Astrophys.\ J.\ Suppl.\  {\bf 148}, 213 (2003).

\bibitem{Page:2003fa}
L. Page {\it et al.},
Astrophys.\ J.\ Suppl.\  {\bf 148}, 233 (2003).

\bibitem{Moodley:2004nz}
K. Moodley, M. Bucher, J. Dunkley, P. G. Ferreira and C. Skordis,
arXiv:astro-ph/0407304.

\bibitem{Bucher:2004an}
M. Bucher, J. Dunkley, P. G. Ferreira, K. Moodley and C. Skordis,
Phys.\ Rev.\ Lett.\  {\bf 93}, 081301 (2004).

\bibitem{Bucher:2000hy}
M. Bucher, K. Moodley and N. Turok,
Phys.\ Rev.\ Lett.\  {\bf 87}, 191301 (2001).

\bibitem{Enqvist:2000hp}
K. Enqvist, H. Kurki-Suonio and J. Valiviita,
Phys.\ Rev.\ D {\bf 62}, 103003 (2000).

\bibitem{Beltran:2004uv}
M. Beltran, J. Garcia-Bellido, J. Lesgourgues and A. Riazuelo,
arXiv:astro-ph/0409326.

\bibitem{Crotty:2003rz}
P. Crotty, J. Garcia-Bellido, J. Lesgourgues and A. Riazuelo,
Phys.\ Rev.\ Lett.\  {\bf 91}, 171301 (2003).

\bibitem{Amendola:2001ni}
L. Amendola, C. Gordon, D. Wands and M. Sasaki,
Phys.\ Rev.\ Lett.\  {\bf 88}, 211302 (2002).

\bibitem{Chung:2001cb}
D. J. H. Chung, P. Crotty, E. W. Kolb and A. Riotto,
Phys.\ Rev.\ D {\bf 64}, 043503 (2001).

\bibitem{Chung:1998zb}
D. J. H. Chung, E. W. Kolb and A. Riotto,
Phys.\ Rev.\ D {\bf 59}, 023501 (1999);
D. J. H. Chung, E. W. Kolb and A. Riotto,
Phys.\ Rev.\ Lett.\  {\bf 81} (1998) 4048;
E. W. Kolb, D. J. H. Chung and A. Riotto,
arXiv:hep-ph/9810361.

\bibitem{Kuzmin:1998uv}
V. Kuzmin and I. Tkachev,
JETP Lett.\  {\bf 68}, 271 (1998).

\bibitem{Kuzmin:1998kk}
V. Kuzmin and I. Tkachev,
Phys.\ Rev.\ D {\bf 59}, 123006 (1999).

\bibitem{Chung:1998bt}
D. J. H. Chung,
Phys.\ Rev.\ D {\bf 67}, 083514 (2003).

\bibitem{Benakli:1998ut}
K. Benakli, J. R. Ellis and D. V. Nanopoulos,
Phys.\ Rev.\ D {\bf 59}, 047301 (1999).

\bibitem{Han:1998pa}
T. Han, T. Yanagida and R. J. Zhang,
Phys.\ Rev.\ D {\bf 58}, 095011 (1998).

\bibitem{Hamaguchi:1998wm}
K. Hamaguchi, Y. Nomura and T. Yanagida,
Phys.\ Rev.\ D {\bf 58}, 103503 (1998).

\bibitem{Benakli:1998sy}
K. Benakli,
Phys.\ Lett.\ B {\bf 447}, 51 (1999).

\bibitem{Leontaris:1999ce}
G. K. Leontaris and J. Rizos,
Nucl.\ Phys.\ B {\bf 554}, 3 (1999).

\bibitem{Hamaguchi:1999cv}
K. Hamaguchi, K. I. Izawa, Y. Nomura and T. Yanagida,
Phys.\ Rev.\ D {\bf 60}, 125009 (1999).

\bibitem{Dvali:1999tq}
G. R. Dvali,
Phys.\ Lett.\ B {\bf 459}, 489 (1999).

\bibitem{Coriano:2001mg}
C. Coriano, A. E. Faraggi and M. Plumacher,
Nucl.\ Phys.\ B {\bf 614}, 233 (2001).

\bibitem{Uehara:2001wd}
Y. Uehara,
JHEP {\bf 0112}, 034 (2001).

\bibitem{Ellis:2004cj}
J. R. Ellis, V. E. Mayes and D. V. Nanopoulos,
arXiv:hep-ph/0403144.

\bibitem{Shiu:2003ta}
G. Shiu and L. T. Wang,
Phys.\ Rev.\ D {\bf 69}, 126007 (2004).

\bibitem{Weinberg:1982zq}
S. Weinberg,
Phys.\ Rev.\ Lett.\  {\bf 48}, 1303 (1982).

\bibitem{Ellis:1984er}
J. R. Ellis, D. V. Nanopoulos and S. Sarkar,
Nucl.\ Phys.\ B {\bf 259}, 175 (1985).

\bibitem{Lindley:1986wt}
D. Lindley,
Phys.\ Lett.\ B {\bf 171}, 235 (1986).

\bibitem{Kawasaki:1986my}
M. Kawasaki and K. Sato,
Phys.\ Lett.\ B {\bf 189}, 23 (1987).

\bibitem{Moroi:1993mb}
T. Moroi, H. Murayama and M. Yamaguchi,
Phys.\ Lett.\ B {\bf 303}, 289 (1993).

\bibitem{Kohri:2001jx}
K. Kohri,
Phys.\ Rev.\ D {\bf 64}, 043515 (2001).

\bibitem{Kawasaki:2004qu}
M. Kawasaki, K. Kohri and T. Moroi,
arXiv:astro-ph/0408426.

\bibitem{Kawasaki:2004fw}
M. Kawasaki, K. Kohri and T. Moroi,
arXiv:hep-ph/0410287.

\bibitem{Chung:2003fi}
D. J. H. Chung, L. L. Everett, G. L. Kane, S. F. King, J. Lykken and L. T. Wang,
arXiv:hep-ph/0312378.

\bibitem{Knox:2002pe}
L. Knox and Y. S. Song,
Phys.\ Rev.\ Lett.\  {\bf 89}, 011303 (2002).

\bibitem{Song:2003ca}
Y. S. Song and L. Knox,
Phys.\ Rev.\ D {\bf 68}, 043518 (2003).

\bibitem{Mukhanov:2003xr}
V. Mukhanov,
arXiv:astro-ph/0303072.

\bibitem{Hu:1994uz}
W. Hu and N. Sugiyama,
Astrophys.\ J.\  {\bf 444}, 489 (1995).

\bibitem{Hu:1994jd}
W. Hu and N. Sugiyama,
Phys.\ Rev.\ D {\bf 51}, 2599 (1995).

\bibitem{Liddle:1999pr}
A. R. Liddle and A. Mazumdar,
Phys.\ Rev.\ D {\bf 61}, 123507 (2000).

\end{thebibliography}
\end{document}